\PassOptionsToPackage{unicode}{hyperref}
\PassOptionsToPackage{hyphens}{url}
\documentclass[
]{article}
\usepackage{amsmath,amssymb}
\usepackage{iftex}
\ifPDFTeX
  \usepackage[T1]{fontenc}
  \usepackage[utf8]{inputenc}
  \usepackage{textcomp} % provide euro and other symbols
\else % if luatex or xetex
  \usepackage{unicode-math} % this also loads fontspec
  \defaultfontfeatures{Scale=MatchLowercase}
  \defaultfontfeatures[\rmfamily]{Ligatures=TeX,Scale=1}
\fi
\usepackage{lmodern}
\ifPDFTeX\else
\fi
\IfFileExists{upquote.sty}{\usepackage{upquote}}{}
\IfFileExists{microtype.sty}{% use microtype if available
  \usepackage[]{microtype}
  \UseMicrotypeSet[protrusion]{basicmath} % disable protrusion for tt fonts
}{}
\makeatletter
\@ifundefined{KOMAClassName}{% if non-KOMA class
  \IfFileExists{parskip.sty}{%
    \usepackage{parskip}
  }{% else
    \setlength{\parindent}{0pt}
    \setlength{\parskip}{6pt plus 2pt minus 1pt}}
}{% if KOMA class
  \KOMAoptions{parskip=half}}
\makeatother
\usepackage{xcolor}
\usepackage[margin=1in]{geometry}
\usepackage{color}
\usepackage{fancyvrb}

\DefineVerbatimEnvironment{Highlighting}{Verbatim}{commandchars=\\\{\}}
\usepackage{framed}
\definecolor{shadecolor}{RGB}{248,248,248}
\newenvironment{Shaded}{\begin{snugshade}}{\end{snugshade}}

\newcommand{\AttributeTok}[1]{\textcolor[rgb]{0.13,0.29,0.53}{#1}}

\newcommand{\CommentTok}[1]{\textcolor[rgb]{0.56,0.35,0.01}{\textit{#1}}}

\newcommand{\ConstantTok}[1]{\textcolor[rgb]{0.56,0.35,0.01}{#1}}
\newcommand{\ControlFlowTok}[1]{\textcolor[rgb]{0.13,0.29,0.53}{\textbf{#1}}}

\newcommand{\DecValTok}[1]{\textcolor[rgb]{0.00,0.00,0.81}{#1}}
\newcommand{\DocumentationTok}[1]{\textcolor[rgb]{0.56,0.35,0.01}{\textbf{\textit{#1}}}}

\newcommand{\FloatTok}[1]{\textcolor[rgb]{0.00,0.00,0.81}{#1}}
\newcommand{\FunctionTok}[1]{\textcolor[rgb]{0.13,0.29,0.53}{\textbf{#1}}}

\newcommand{\NormalTok}[1]{#1}

\newcommand{\OtherTok}[1]{\textcolor[rgb]{0.56,0.35,0.01}{#1}}

\newcommand{\SpecialCharTok}[1]{\textcolor[rgb]{0.81,0.36,0.00}{\textbf{#1}}}

\newcommand{\StringTok}[1]{\textcolor[rgb]{0.31,0.60,0.02}{#1}}

\usepackage{longtable,booktabs,array}
\usepackage{calc} % for calculating minipage widths
\usepackage{etoolbox}
\makeatletter
\patchcmd\longtable{\par}{\if@noskipsec\mbox{}\fi\par}{}{}
\makeatother
\IfFileExists{footnotehyper.sty}{\usepackage{footnotehyper}}{\usepackage{footnote}}
\makesavenoteenv{longtable}
\usepackage{graphicx}
\makeatletter
\def\maxwidth{\ifdim\Gin@nat@width>\linewidth\linewidth\else\Gin@nat@width\fi}
\def\maxheight{\ifdim\Gin@nat@height>\textheight\textheight\else\Gin@nat@height\fi}
\makeatother
\setkeys{Gin}{width=\maxwidth,height=\maxheight,keepaspectratio}
\makeatletter
\def\fps@figure{htbp}
\makeatother
\providecommand{\tightlist}{%
  \setlength{\itemsep}{0pt}\setlength{\parskip}{0pt}}
\newlength{\cslhangindent}
\newlength{\csllabelwidth}
\newlength{\cslentryspacingunit} % times entry-spacing
\newenvironment{CSLReferences}[2] % #1 hanging-ident, #2 entry spacing
 {% don't indent paragraphs
  \setlength{\parindent}{0pt}
  \ifodd #1
  \let\oldpar\par
  \def\par{\hangindent=\cslhangindent\oldpar}
  \fi
  \setlength{\parskip}{#2\cslentryspacingunit}
 }%
 {}
\usepackage{calc}

\newcommand{\CSLLeftMargin}[1]{\parbox[t]{\csllabelwidth}{#1}}
\newcommand{\CSLRightInline}[1]{\parbox[t]{\linewidth - \csllabelwidth}{#1}\break}

\usepackage[disable]{endfloat}
\usepackage{booktabs}
\usepackage{longtable}
\usepackage{array}
\usepackage{multirow}
\usepackage{wrapfig}
\usepackage{float}
\usepackage{colortbl}
\usepackage{pdflscape}
\usepackage{tabu}
\usepackage{threeparttable}
\usepackage{threeparttablex}
\usepackage[normalem]{ulem}
\usepackage{makecell}
\usepackage{xcolor}
\ifLuaTeX
  \usepackage{selnolig}  % disable illegal ligatures
\fi
\IfFileExists{bookmark.sty}{\usepackage{bookmark}}{\usepackage{hyperref}}
\IfFileExists{xurl.sty}{\usepackage{xurl}}{} % add URL line breaks if available
\hypersetup{
  pdftitle={Standardised workflow for mass spectrometry-based single-cell proteomics data processing and analysis using the scp package},
  pdfauthor={Samuel Grégoire; Christophe Vanderaa\^{}\{*\}; Sébastien Pyr dit Ruys; Christopher Kune; Gabriel Mazzucchelli ; Didier Vertommen; Laurent Gatto\^{}\{*\}},
  hidelinks,
  pdfcreator={LaTeX via pandoc}}

\title{Standardised workflow for mass spectrometry-based single-cell proteomics data processing and analysis using the \texttt{scp} package}
\author{Samuel Grégoire\footnote{Computational Biology and Bioinformatics Unit, de Duve Institute, UCLouvain, Brussels, Belgium} \and Christophe Vanderaa\(^{*}\) \and Sébastien Pyr dit Ruys\footnote{Protein Phosphorylation Unit, de Duve Institute, UCLouvain, Brussels, Belgium} \and Christopher Kune\footnote{Laboratory of Mass Spectrometry, MolSys Research Unit, University of Liège, Belgium} \and Gabriel Mazzucchelli\footnote{GIGA Proteomics Facility, University of Liège, Belgium} \textsuperscript{\textdaggerdbl} \and Didier Vertommen\textsuperscript{\textdagger} \and Laurent Gatto\(^{*}\)\footnote{\href{mailto:laurent.gatto@uclouvain.be}{\nolinkurl{laurent.gatto@uclouvain.be}}}}
\date{}

\begin{document}
\maketitle

\textbf{Abstract} Mass spectrometry (MS) based single-cell proteomics (SCP)
explores cellular heterogeneity by focusing on the functional
effectors of the cells - proteins. However, extracting meaningful
biological information from MS data is far from trivial, especially
with single cells. Currently, data analysis workflows are
substantially different from one research team to another. Moreover,
it is difficult to evaluate pipelines as ground truths are missing.
Our team has developed the R/Bioconductor package called \texttt{scp} to
provide a standardised framework for SCP data analysis. It relies on
the widely used \texttt{QFeatures} and \texttt{SingleCellExperiment} data
structures. In addition, we used a design containing cell lines mixed
in known proportions to generate controlled variability for data
analysis benchmarking. In this work, we provide a flexible data
analysis protocol for SCP data using the \texttt{scp} package together with
comprehensive explanations at each step of the processing. Our main
steps are quality control on the feature and cell level, aggregation
of the raw data into peptides and proteins, normalisation and batch
correction. We validate our workflow using our ground truth data
set. We illustrate how to use this modular, standardised framework and
highlight some crucial steps.

\bigskip


\bigskip

\hypertarget{introduction}{%
\section{Introduction}\label{introduction}}

Single-cell proteomics (SCP) aims at studying cellular heterogeneity
by focusing on the functional effectors of the cells - proteins. Mass
spectrometry (MS) has been established as the method of choice for
exploring the proteome, and has logically expanded into single-cell
proteomics. Recent breakthroughs in instrument performances and both
label-free and multiplexed fields\textsuperscript{\protect\hyperlink{ref-leduc_exploring_2022}{1}--\protect\hyperlink{ref-matzinger_robust_2023}{3}} opened perspectives
for practical application of SCP as a tool to study cell physiology,
cancer development or drug resistance, among others\textsuperscript{\protect\hyperlink{ref-slavov_learning_2022}{4}}.

However, extracting meaningful biological information from the complex
data generated with mass spectrometry is far from trivial, especially
when working with single cells. With the development of SCP comes the
need for suitable data processing workflows. Currently, most research
teams rely on custom scripts and software to analyse their data. This
implies that each team has their own workflow, with a wide variety of
steps, each impacting the outcome of the processing\textsuperscript{\protect\hyperlink{ref-vanderaa_current_2023}{5},\protect\hyperlink{ref-vanderaa_replication_2021}{6}}. The development
of standardised tools for SCP data analysis unifies existing workflows
and, hence, facilitates the access and the spreading of SCP analysis
to other labs while improving reproducibility. With this in mind, our
team has developed an R/Bioconductor package called \texttt{scp}\textsuperscript{\protect\hyperlink{ref-vanderaa_replication_2021}{6}} to provide a standardised framework for
SCP data analysis. The \texttt{scp} package is designed as a modular tool
where each processing step returns a consistent and standardised
output that can easily be chained into the next one. Therefore, steps
can be arranged in different ways to build and test workflows. The
software is part of the Bioconductor project\textsuperscript{\protect\hyperlink{ref-huber_orchestrating_2015}{7}}. The project is well known for its
exemplary coding practices, effortless interoperability between its
software packages, thoroughly maintained and centralised
infrastructure, and its commitment to reproducibility and long-term
maintenance.

\begin{figure}

{\centering \includegraphics[width=440px]{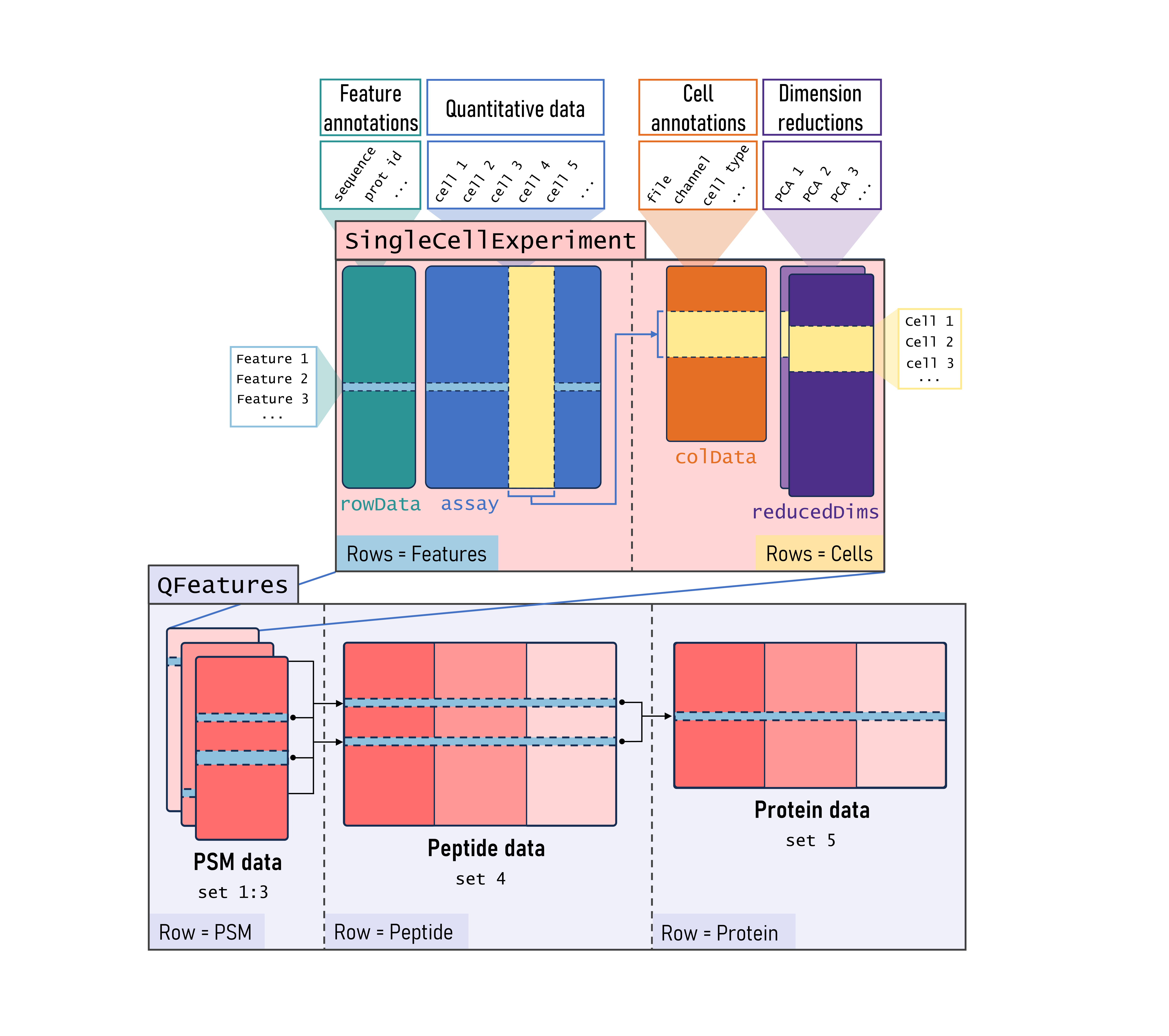} 

}

\caption{\texttt{SingleCellExperiment} and \texttt{QFeatures} classes and the \texttt{scp} framework. Figure adapted from Amezquita et al. 2020. Quantitative data are stored inside \texttt{SingleCellExperiment} objects alongside feature annotations, cell annotations and dimension-reduced data. Rows of the feature annotations match rows of the quantitative data, and rows of the cell annotations and dimension-reduced data match the columns of the quantitative data, i.e. the single cells. \texttt{SingleCellExperiment} objects are stored together inside a \texttt{QFeatures} object. The \texttt{QFeatures} object typically contains several \texttt{SingleCellExperiment} objects corresponding to the PSM sets (one for each MS run), the joined peptide data and finally, the protein data.}\label{fig:seqf}
\end{figure}

A data structure, also known as a data class, is a specialised format
designed for storing, organising, retrieving and processing data. We
will use the term ``data object'' to refer to data that adheres to a
given data structure. The \texttt{scp} framework relies on two data
structures. The first one is the \texttt{SingleCellExperiment}\textsuperscript{\protect\hyperlink{ref-R-SingleCellExperiment}{8},\protect\hyperlink{ref-SingleCellExperiment2020}{9}} class. It stores
the different pieces of data collected during a single-cell
experiment, such as the measured quantities or the cell and feature
annotations, to facilitate their simultaneous manipulation (\textbf{Figure
\ref{fig:seqf}}, top). The quantitative data is typically a table
output by the pre-processing software with label-free or TMT channel
intensities for each identified spectrum. It is stored in a
quantification matrix called \texttt{assay} with samples (single cells in
this case) aligned along columns and features aligned along rows.
Features can be any type of measurable biological entity. In
particular, MS-based proteomics deals with peptide spectrum matches
(PSM) where recorded MS spectra can be assigned to peptide
sequences. Feature annotations (rows) make up the \texttt{rowData}
slot. Feature annotations are supplementary information generated by
the pre-processing software like peptide sequence, protein name and
ion charge. Since the \texttt{rowData} slot contains the features
annotations, its rows are associated with the rows of the quantitative
\texttt{assay}. Cell (column) annotations make up the \texttt{colData} slot. The
table is provided by the experimenter and documents potential sources
of biological or technical variation, such as cell type or acquisition
batch. Each row of the cell annotation table represents a single cell
while its attributes are defined along the columns of the table. The
\texttt{SingleCellExperiment} structure serves as an interface for a wide
range of packages dealing with single-cell data analysis and plays
therefore a key role in ensuring compatibility between single-cell
methods across different fields.

The second data structure used by the \texttt{scp} package is the \texttt{QFeatures}
class. \texttt{QFeatures} is designed for managing and processing the
quantitative features from high-throughput MS experiments. \texttt{QFeatures}
provides access to many generic approaches for MS-based proteomics
analysis. It can store multiple \texttt{SingleCellExperiment} objects (which
we will call ``sets'' below) while preserving the hierarchical
relationship between features from different sets (\textbf{Figure
\ref{fig:seqf}}, bottom). The hierarchical data structure enabled by
\texttt{QFeatures} is of particular interest for MS-based proteomics data as
proteins are composed of peptides, themselves inferred from PSMs. Sets
can be joined and manipulated and their relations are tracked and
recorded, thus allowing users to easily navigate across PSM, peptide
and protein quantitative data. In short, the package \texttt{scp} manages
\texttt{SingleCellExperiment} inside \texttt{QFeatures}.

To evaluate the data processing steps and refine our workflow, we
generated a SCP benchmarking dataset. We used a design containing cell
lines mixed in known proportions to generate controlled
variability. Mixture designs generate data that exhibit biological
heterogeneity with available ground truth. They have been successfully
applied to single-cell RNA-seq\textsuperscript{\protect\hyperlink{ref-tian_benchmarking_2019}{10},\protect\hyperlink{ref-Mereu2020-vk}{11}}. In addition, we added a second layer of heterogeneity
by including differentiating cells. We induced differentiation of
both U937 and THP1 cell lines to emulate the complexity of a
biological sample, hence generating a data set that is more closely
related to real-life applications.

\begin{figure}

\includegraphics[width=450px]{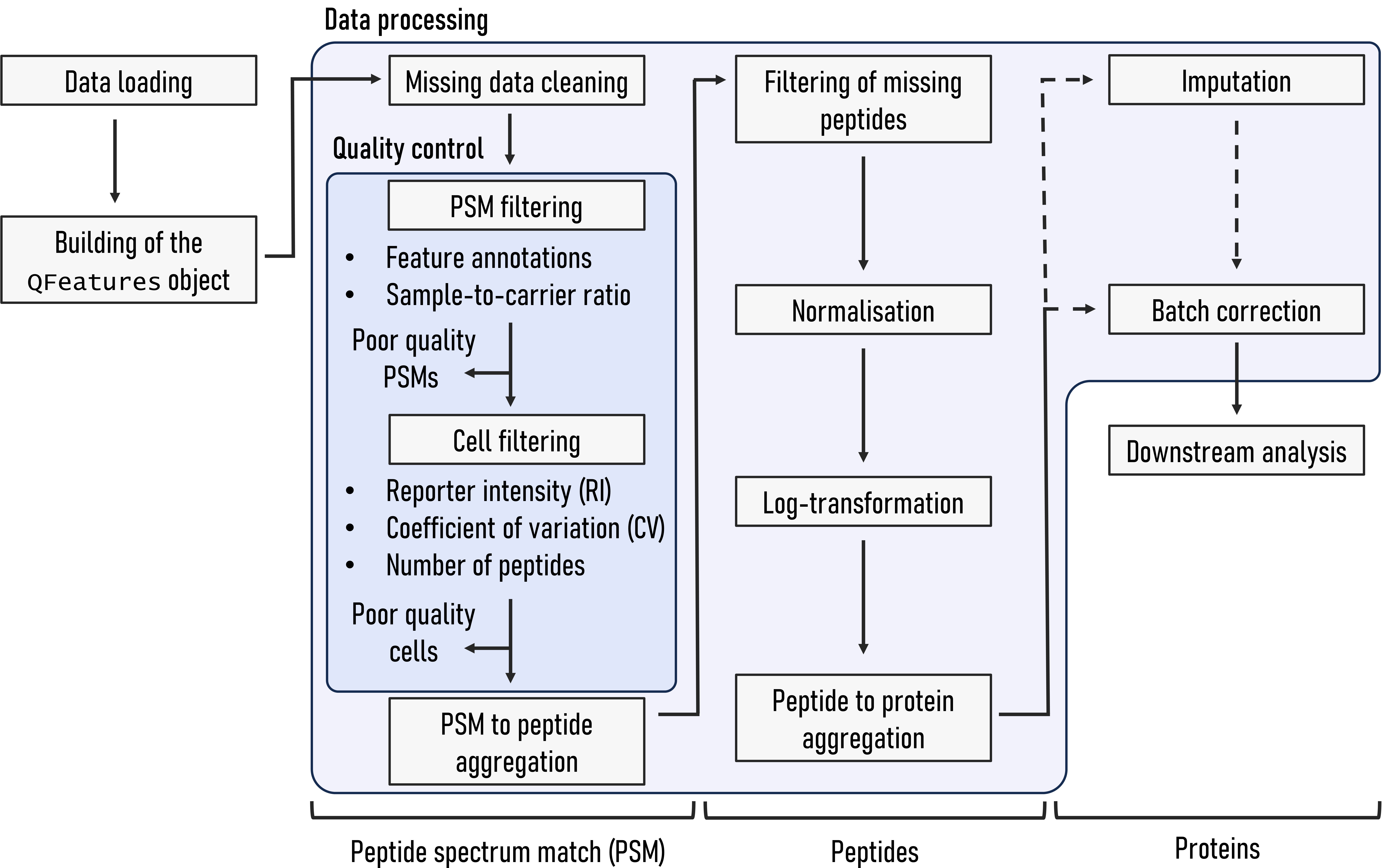} \hfill{}

\caption{MS-based single-cell proteomics data analysis workflow. We load the data and build a \texttt{QFeatures} object before proceeding to data processing. We then format the missing values to appear as missing values (NA) and not as 0. Quality control is performed at two levels: the features, here PSMs, and the samples, here single cells. PSM filtering uses features annotations and the sample-to-carrier ratio metric to remove poor quality PSMs. Cell filtering uses 3 metrics, reporter intensity (RI), coefficient of variation (CV) and the number of peptides to remove poor quality cells. The PSM data is aggregated into the peptides data which is normalised, log-transformed and finally aggregated into the protein data. Finaly, we apply batch correction to the protein data. Processed data are ready for downstream analysis. Imputation is optional (dashed arrows).}\label{fig:wf}
\end{figure}

In this chapter, we describe typical SCP data processing using the
\texttt{scp} package, as illustrated schematically on \textbf{Figure
\ref{fig:wf}}. Note that all of these steps are demonstrated with
carrier-based TMTpro multiplexed samples, acquired with data-dependent
acquisition (DDA) mode. Unless stated otherwise, the steps in his
protocol are applicable to other data types, including data acquired
using different multiplexing reagents (e.g., mTRAQ), data acquired for
label-free samples, data containing any number of cells, or data
acquired with both DDA and data-independent acquisition (DIA) mode.

In the following sections, we will describe how data is loaded to
build a \texttt{QFeature} data object, and then proceed to quantitative data
processing. Our main steps consist of: quality control, aggregation
into peptides and proteins, normalisation and batch correction. We
will conclude with dimensionality reduction on the resulting protein
data.

\hypertarget{materials}{%
\section{Materials}\label{materials}}

\hypertarget{installation}{%
\subsection{Installation}\label{installation}}

The analyses presented below require several Bioconductor packages. To
install Bioconductor packages you need to install the \texttt{BiocManager}
package from the Comprehensive R Archive Network (CRAN) by running
\texttt{install.packages("BiocManager")}. Packages that are directly used in
this workflow are listed below. A complete list of required packages
is available in the \emph{Session information} section.

\begin{itemize}
\tightlist
\item
  \texttt{QFeatures} is used for manipulation of \texttt{QFeature} data structure
  and filtering.
\item
  \texttt{scp} is used to build a \texttt{QFeatures} object with SCP data\textsuperscript{\protect\hyperlink{ref-vanderaa_replication_2021}{6}}. It also provides functions to compute
  the sample-to-carrier ratio and to compute the median coefficient of variation
  per cell.
\item
  \texttt{dplyr} is used for basic data manipulation with functions like
  \texttt{filter()}\textsuperscript{\protect\hyperlink{ref-Hadley_dplyr_2023}{12}}.
\item
  \texttt{ggplot2} and \texttt{patchwork} are used for visualisation\textsuperscript{\protect\hyperlink{ref-Hadley_gpglot2_2016}{13}}.
\item
  \texttt{limma} is used for batch correction\textsuperscript{\protect\hyperlink{ref-Ritchie_limma_2015}{14}}.
\item
  \texttt{scater} is used for dimensionality reduction and visualisation of
  reduced dimensions\textsuperscript{\protect\hyperlink{ref-McCarthy_scater_2017}{15}}.
\end{itemize}

All packages can be installed in the same way by running
\texttt{BiocManager::install("package\_name")}. For example, run
\texttt{BiocManager::install("scp")} to install \texttt{scp} and all its
dependencies.

\hypertarget{dataset}{%
\subsection{Dataset}\label{dataset}}

The dataset used to illustrate SCP data analysis has been generated
in-house or through collaborations and is distributed across 8
acquisition batches (\textbf{Table \ref{tab:datatab}}). Samples were
prepared using the SCoPE2 protocol\textsuperscript{\protect\hyperlink{ref-specht_single-cell_2021}{16}}, as
described in section \ref{datagen}. We here provide a brief
description of the experimental design. Cells come from THP1 and U937
human monocyte cell lines. Samples are either single cells (sc) or
single-cell equivalents (sc\_eq) i.e.~peptides extracted from bulk
samples but diluted to single-cell range (0.3 ng). For batches CBIO681
and CBIO703, single-cell equivalents from THP1 and U937 were combined
to create ``Mix'' equivalents. For CBIO715, ``Mix'' samples are generated
by sorting one THP1 and one U937 cell in the same well. All samples
were run on the Orbitrap Fusion Lumos Tribrid™ mass spectrometer
except for samples from batch CBIO733 and GIGA. The batch CBIO733 was
run in-house on the Orbitrap Exploris 240. The batch GIGA was run at
the GIGA institute from ULiège on the Brucker timsTOF SCP.

\begin{table}

\caption{\label{tab:datatab}Description of the 8 acquistion batches of the SCP dataset. Rows represent acquisition batches. The columns describe their names, the cell types used, the number of single cells analysed, the type of sample, i.e. single cells (sc) or single-cell equivalents (sc\_eq), and the type of MS instrument used.}
\centering
\begin{tabular}[t]{lllrll}
\toprule
  & Batch name & Cell type & Single cells & Sample type & Mass spectrometer\\
\midrule
1 & CBIO680 & THP1, U937 & 12 & sc\_eq & Lumos Tribrid\\
2 & CBIO681 & THP1, U937, Mix & 40 & sc\_eq & Lumos Tribrid\\
3 & CBIO703 & THP1, U937, Mix & 40 & sc\_eq & Lumos Tribrid\\
4 & CBIO715 & THP1, U937, Mix & 68 & sc & Lumos Tribrid\\
5 & CBIO725 & THP1, U937, THP1\_dif, U937\_dif & 120 & sc & Lumos Tribrid\\
\addlinespace
6 & CBIO733 & THP1, U937, THP1\_dif, U937\_dif & 120 & sc & Exploris 240\\
7 & CBIO754 & THP1, U937, THP1\_dif, U937\_dif & 120 & sc & Lumos Tribrid\\
8 & GIGA & THP1, U937 & 36 & sc & timsTOF SCP\\
\bottomrule
\end{tabular}
\end{table}

\hypertarget{methods}{%
\section{Methods}\label{methods}}

\hypertarget{datagen}{%
\subsection{Data generation}\label{datagen}}

SCP samples were prepared using the SCoPE2 protocol\textsuperscript{\protect\hyperlink{ref-specht_single-cell_2021}{16}}. In short, single cells were isolated from
THP1 and U937 cell lines in 384-well plates using the BD FACSAria™ III
Cell Sorter. Cells were lysed using a hypotonic shock followed by a
thermic shock and sonication. Single-cell equivalents were generated
by diluting bulk THP1 and U937 lysates to dispense 0.3 ng of lysate
per well. THP1 and U937 differentiation was induced by 48h treatment
with phorbol 12-myristate-13-acetate (PMA). Lysates were digested
using trypsin and peptides were labelled with TMTpro 16-plex
labels. For the GIGA acquisition batch, sets were labelled with either
126C, 127N, 128C, 129N, 130C, 131N, 132C, 133N or 127C, 128N, 129C,
130N, 131C, 132N, 133C, 134N and are \emph{de facto} 8 plex samples. This
was done to account for the lower resolution of the timsTOF SCP, which
cannot distinguish between C and N labels of the same mass. Labelled
samples were pooled with a labelled carrier sample containing peptides
from 50 cells, and injected into the Ultimate 3000 LC System (CBIO
batches) or the Vanquish™ Neo UHPLC System (GIGA batch) for liquid
chromatography (LC). BioZen™ Peptide Polar C18 250 x 0.0075mm columns
were used for LC with a 120 minute gradient (CBIO batches) or a 30
minute gradient (GIGA batches). Samples were run on either the
Orbitrap Fusion Lumos Tribrid™, the Exploris 240 or the timsTOF SCP
mass spectrometer (\textbf{Table \ref{tab:datatab}}).

\hypertarget{data-preprocessing}{%
\subsection{Data preprocessing}\label{data-preprocessing}}

Raw data files were converted into mzML format using the \texttt{MSconvert}
software\textsuperscript{\protect\hyperlink{ref-adusumilli_data_2017}{17}}. mzML files were searched by the
\texttt{sage} software\textsuperscript{\protect\hyperlink{ref-lazear_sage_2023}{18}} against a protein sequence
database including all entries from the human SwissProt database
(downloaded December 23, 2022). The \texttt{results.json} configuration file
can be found in our Zenodo repository. In short, we specified cleaving
sites as lysine and arginine, allowed for 2 missed cleavages and
limited the search to peptides ranging between 5 to 50 amino
acids. Cysteine carbamidomethylation was the only fixed modification
and lysine TMTpro, peptide N-terminal TMTpro, methionine oxydation and
protein N-terminal acetylation were set as variable
modifications. Quantitative and identification results were exported
and merged, as documented in the \texttt{build\_QF\_dataset.Rmd} file. All
these files are available in a Zenodo archive (10.5281/zenodo.8417228)\textsuperscript{\protect\hyperlink{ref-zenododata}{19}}

\hypertarget{data-availability}{%
\subsection{Data availability}\label{data-availability}}

The \texttt{.raw} and \texttt{.mzML} files, and \texttt{d} folders, the \texttt{sage} results
files and the complete R-ready data, i.e.~all files required for
running this protocol, are available in a Zenodo archive
(10.5281/zenodo.8417228)\textsuperscript{\protect\hyperlink{ref-zenododata}{19}}. Analyses can be fully
reproduced by loading the SCP data from the \texttt{scp.rds} file and running
the code presented in the following sections. Construction of the data
in the \texttt{scp.rds} file can be reproduced using the search engine
outputs, the cell annotations and the \texttt{build\_QF\_dataset.Rmd} R script,
also available in the Zenodo archive. The raw mass spectrometry data
have also been deposited to the ProteomeXchange Consortium\textsuperscript{\protect\hyperlink{ref-Vizcaino2014}{20}} via the PRIDE partner repository with the dataset
identifier PXD046211.

\hypertarget{packages-and-data-loading}{%
\subsection{Packages and data loading}\label{packages-and-data-loading}}

The functions required for executing the protocol are only available
when the packages are loaded. Packages should be loaded using the
function \texttt{library()} every time a new R session is opened. Note that
the \texttt{QFeatures} package is automatically loaded with \texttt{scp}.

\begin{Shaded}
\begin{Highlighting}[]
\FunctionTok{library}\NormalTok{(}\StringTok{"scp"}\NormalTok{)}
\FunctionTok{library}\NormalTok{(}\StringTok{"dplyr"}\NormalTok{)}
\FunctionTok{library}\NormalTok{(}\StringTok{"ggplot2"}\NormalTok{)}
\FunctionTok{library}\NormalTok{(}\StringTok{"limma"}\NormalTok{)}
\FunctionTok{library}\NormalTok{(}\StringTok{"scater"}\NormalTok{)}
\FunctionTok{library}\NormalTok{(}\StringTok{"patchwork"}\NormalTok{)}
\end{Highlighting}
\end{Shaded}

To build the \texttt{QFeatures} object, we need two particular tables:

\begin{itemize}
\item
  The quantitative input table containing the features (typically
  PSMs) quantification, acquisition annotations such as the file name,
  and feature annotations such as peptide sequence, ion charge and
  protein name. It is generated by a pre-processing software such as
  \texttt{MaxQuant}\textsuperscript{\protect\hyperlink{ref-cox_maxquant_2008}{21}}, \texttt{ProteomeDiscoverer} (Thermo Fisher
  Scientific), \texttt{MSFragger}\textsuperscript{\protect\hyperlink{ref-kong_msfragger_2017}{22}}, or \texttt{sage}\textsuperscript{\protect\hyperlink{ref-lazear_sage_2023}{18}}. In this protocol, we merge the quantitation and
  identification result files \texttt{quant.tsv} and \texttt{results.sage.tsv}
  generated by \texttt{sage} to create the input table.
\item
  The sample table containing the experimental design generated by the
  researcher. The experimental design should be reported as a table
  listing each analyzed single cell along the rows and each descriptor
  along the columns. Descriptors can be either biological, such as the
  sample type, the cell type or the patient identifier, or technical,
  such as raw data file names and date of acquisition. When available,
  additional descriptors, biological or technical, should be provided
  to perform quality control or help explain a specific pattern in the
  data. More information about those descriptors can be found in
  recent recommendations and guidelines for single-cell proteomics
  experiments\textsuperscript{\protect\hyperlink{ref-gatto_initial_2023}{23}}.
\end{itemize}

\hypertarget{quanttab}{%
\subsubsection{Quantitative input table}\label{quanttab}}

\begin{Shaded}
\begin{Highlighting}[]
\NormalTok{scp\_subset }\OtherTok{\textless{}{-}} \FunctionTok{read.csv}\NormalTok{(}\StringTok{"./data/scp\_subset.csv"}\NormalTok{, }\AttributeTok{check.names =} \ConstantTok{FALSE}\NormalTok{)}

\NormalTok{scp\_subset}
\end{Highlighting}
\end{Shaded}

\begin{verbatim}
##           run      128N     128C     129N     129C             peptide
## 1  CBIO725_10 6922.1846 5263.048 7098.995 5040.178       LPLQQTTFPHQLR
## 2  CBIO725_10 2095.7988 3408.995 4171.487 4699.922              IHGTFK
## 3  CBIO725_10    0.0000    0.000    0.000    0.000 GRRTGSPGEGAHVSAAVAK
## 4  CBIO725_10 6075.7397 7903.320 6749.178 4174.352             RGIFDDR
## 5  CBIO725_10 6188.5337 5289.025 8552.856 9242.520            LSYSLKKR
## 6  CBIO754_16 1612.5588 3163.930 2093.486 1091.929      SVIQRLPSIDCIVR
## 7  CBIO754_16  899.0344 6366.098 1993.906 1321.362              DLVFKR
## 8  CBIO754_16    0.0000 3342.914    0.000 1370.106          SADTLWDIQK
## 9  CBIO754_16 2844.2880 2871.017 1694.564 2586.150       TLNDELEIIEGMK
## 10 CBIO754_16    0.0000 1357.731    0.000    0.000      KEETFALYRDVWMK
##              proteins  peptide_fdr
## 1   Q12851|M4K2_HUMAN 0.7661977400
## 2   Q15542|TAF5_HUMAN 0.6765558000
## 3    P30518|V2R_HUMAN 0.8892745000
## 4  Q6DN14|MCTP1_HUMAN 0.7640124000
## 5  B1ANY3|F220P_HUMAN 0.6833753600
## 6  Q9UGU0|TCF20_HUMAN 0.9123871000
## 7    P20711|DDC_HUMAN 0.2462402100
## 8   P07195|LDHB_HUMAN 0.0001707067
## 9   P10809|CH60_HUMAN 0.0001707067
## 10 O14497|ARI1A_HUMAN 0.9042809600
\end{verbatim}

Note that this is only a small subset of the data. It offers a preview
of some of the information that can be found within the quantification
table. The quantitative data that is displayed is restricted to only 4
of the TMT channels (\texttt{128N}, \texttt{128C}, \texttt{129N}, \texttt{129C}) for legibility. A
full dataset is used for data processing in the rest of the protocol.

\hypertarget{sampletab}{%
\subsubsection{Sample table}\label{sampletab}}

\begin{Shaded}
\begin{Highlighting}[]
\NormalTok{coldata\_subset }\OtherTok{\textless{}{-}} \FunctionTok{read.csv}\NormalTok{(}\StringTok{"./data/coldata\_subset.csv"}\NormalTok{, }\AttributeTok{row.names =} \StringTok{"X"}\NormalTok{)}

\NormalTok{coldata\_subset}
\end{Highlighting}
\end{Shaded}

\begin{verbatim}
##                        run channel cell_type sample_type   batch
## CBIO725_10_128N CBIO725_10    128N  THP1_dif        SCeq CBIO725
## CBIO725_10_128C CBIO725_10    128C  THP1_dif        SCeq CBIO725
## CBIO725_10_129N CBIO725_10    129N      THP1        SCeq CBIO725
## CBIO725_10_129C CBIO725_10    129C      U937        SCeq CBIO725
## CBIO754_16_128N CBIO754_16    128N     blank          SC CBIO754
## CBIO754_16_128C CBIO754_16    128C      THP1          SC CBIO754
## CBIO754_16_129N CBIO754_16    129N  U937_dif          SC CBIO754
## CBIO754_16_129C CBIO754_16    129C     blank          SC CBIO754
\end{verbatim}

This table shows the cell annotations associated with the 4 TMT
channels and the 2 runs shown in the example subset.

\hypertarget{building-of-the-qfeatures-object.}{%
\subsubsection{\texorpdfstring{Building of the \texttt{QFeatures} object.}{Building of the QFeatures object.}}\label{building-of-the-qfeatures-object.}}

The sample table and the input table are converted into a \texttt{QFeatures}
object with the \texttt{readSCP()} function. To correctly match the
information from the two tables, the function requires 2 specific
fields in the sample table:

\begin{itemize}
\tightlist
\item
  The first field provides the names of the quantification columns in
  the feature data. In this case, the sample table contains a
  \texttt{channel} column that links to the columns that hold the
  quantitative data in the input table (\texttt{128N}, \texttt{128C}, \texttt{129N},
  \texttt{129C}). An issue with the input table is that each quantitative
  column contains information from multiple MS runs, hence from
  multiple cells. Therefore, \texttt{scp} splits the input table into
  separate tables, one for each MS run.
\item
  The second field provides the names of the acquisition runs. This
  field is used to match each row in the sample table with the
  corresponding split of the input table. In this case, the \texttt{run}
  column, present in both the input table and the sample table, allows
  linking the tables. Note that concatenating of \texttt{run} and \texttt{channel}
  generates unique cell identifiers (see row names of the sample
  table).
\end{itemize}

Hence, the two columns allow \texttt{scp} to correctly split the quantitative
input table and match data that were acquired across multiple
acquisitions.

\begin{Shaded}
\begin{Highlighting}[]
\NormalTok{(scp\_subset }\OtherTok{\textless{}{-}} \FunctionTok{readSCP}\NormalTok{(}\AttributeTok{featureData =}\NormalTok{ scp\_subset,}
                       \AttributeTok{colData =}\NormalTok{ coldata\_subset,}
                       \AttributeTok{batchCol =} \StringTok{"run"}\NormalTok{,}
                       \AttributeTok{channelCol =} \StringTok{"channel"}\NormalTok{))}
\end{Highlighting}
\end{Shaded}

\begin{verbatim}
## An instance of class QFeatures containing 2 assays:
##  [1] CBIO725_10: SingleCellExperiment with 5 rows and 4 columns 
##  [2] CBIO754_16: SingleCellExperiment with 5 rows and 4 columns
\end{verbatim}

The object returned by the \texttt{readSCP()} function is a \texttt{QFeatures}
object containing 2 \texttt{SingleCellExperiment} sets named after the 2 MS
runs. Data are split into two sets where each line represents a
unique PSM and each column represents a unique cell. The input table
contains 5 rows for each run. Therefore, each set of the \texttt{QFeature}
object contains 5 rows. The 4 columns correspond to the 4 TMT channels
quantification shown in section \ref{quanttab}.

\hypertarget{exploring-the-qfeatures-object}{%
\subsubsection{\texorpdfstring{Exploring the \texttt{QFeatures} object}{Exploring the QFeatures object}}\label{exploring-the-qfeatures-object}}

Individual sets can be accessed using double brackets. A set can be
selected using either the index number or the name of the set.

\begin{Shaded}
\begin{Highlighting}[]
\NormalTok{scp\_subset[[}\StringTok{"CBIO725\_10"}\NormalTok{]]  }\DocumentationTok{\#\# Same as scp\_subset[[1]]}
\end{Highlighting}
\end{Shaded}

\begin{verbatim}
## class: SingleCellExperiment 
## dim: 5 4 
## metadata(0):
## assays(1): ''
## rownames(5): PSM1 PSM2 PSM3 PSM4 PSM5
## rowData names(4): run peptide proteins peptide_fdr
## colnames(4): CBIO725_10128N CBIO725_10128C CBIO725_10129N
##   CBIO725_10129C
## colData names(0):
## reducedDimNames(0):
## mainExpName: NULL
## altExpNames(0):
\end{verbatim}

For each set, the quantitative data matrix can be extracted with the
\texttt{assay()} accessor function.

\begin{Shaded}
\begin{Highlighting}[]
\FunctionTok{assay}\NormalTok{(scp\_subset[[}\StringTok{"CBIO725\_10"}\NormalTok{]])}
\end{Highlighting}
\end{Shaded}

\begin{verbatim}
##      CBIO725_10128N CBIO725_10128C CBIO725_10129N CBIO725_10129C
## PSM1       6922.185       5263.048       7098.995       5040.178
## PSM2       2095.799       3408.995       4171.487       4699.922
## PSM3          0.000          0.000          0.000          0.000
## PSM4       6075.740       7903.320       6749.178       4174.352
## PSM5       6188.534       5289.025       8552.856       9242.520
\end{verbatim}

Features (i.e.~PSMs, peptides or proteins) information can be
extracted with the \texttt{rowData()} accessor function.

\begin{Shaded}
\begin{Highlighting}[]
\FunctionTok{rowData}\NormalTok{(scp\_subset[[}\StringTok{"CBIO725\_10"}\NormalTok{]])}
\end{Highlighting}
\end{Shaded}

\begin{verbatim}
## DataFrame with 5 rows and 4 columns
##              run       peptide      proteins peptide_fdr
##      <character>   <character>   <character>   <numeric>
## PSM1  CBIO725_10 LPLQQTTFPH... Q12851|M4K...    0.766198
## PSM2  CBIO725_10        IHGTFK Q15542|TAF...    0.676556
## PSM3  CBIO725_10 GRRTGSPGEG... P30518|V2R...    0.889274
## PSM4  CBIO725_10       RGIFDDR Q6DN14|MCT...    0.764012
## PSM5  CBIO725_10      LSYSLKKR B1ANY3|F22...    0.683375
\end{verbatim}

Cell annotations can be accessed with the \texttt{colData()} accessor
function. For the \texttt{colData()}, double brackets subsetting is not
required since the \texttt{QFeatures} object centrally manages samples across
all sets.

\begin{Shaded}
\begin{Highlighting}[]
\FunctionTok{colData}\NormalTok{(scp\_subset)}
\end{Highlighting}
\end{Shaded}

\begin{verbatim}
## DataFrame with 8 rows and 5 columns
##                        run     channel   cell_type sample_type       batch
##                <character> <character> <character> <character> <character>
## CBIO725_10128N  CBIO725_10        128N    THP1_dif        SCeq     CBIO725
## CBIO725_10128C  CBIO725_10        128C    THP1_dif        SCeq     CBIO725
## CBIO725_10129N  CBIO725_10        129N        THP1        SCeq     CBIO725
## CBIO725_10129C  CBIO725_10        129C        U937        SCeq     CBIO725
## CBIO754_16128N  CBIO754_16        128N       blank          SC     CBIO754
## CBIO754_16128C  CBIO754_16        128C        THP1          SC     CBIO754
## CBIO754_16129N  CBIO754_16        129N    U937_dif          SC     CBIO754
## CBIO754_16129C  CBIO754_16        129C       blank          SC     CBIO754
\end{verbatim}

An individual cell annotation field is accessible through the \texttt{\$}
operator.

\begin{Shaded}
\begin{Highlighting}[]
\NormalTok{scp\_subset}\SpecialCharTok{$}\NormalTok{cell\_type  }\DocumentationTok{\#\# Same as colData(scp\_subset)$cell\_type}
\end{Highlighting}
\end{Shaded}

\begin{verbatim}
## [1] "THP1_dif" "THP1_dif" "THP1"     "U937"     "blank"    "THP1"     "U937_dif"
## [8] "blank"
\end{verbatim}

This \texttt{QFeatures} object contains only a small subset of data and is
only used as an illustrative example. For the following processing,
the full dataset will be used (\textbf{Table \ref{tab:datatab}}). The
complete dataset can be readily downloaded as a \texttt{QFeatures} object and
loaded using the \texttt{readRDS()} function.

\begin{Shaded}
\begin{Highlighting}[]
\NormalTok{scp }\OtherTok{\textless{}{-}} \FunctionTok{readRDS}\NormalTok{(}\StringTok{"./data/scp.rds"}\NormalTok{)}
\end{Highlighting}
\end{Shaded}

The dataset contains data for 4 different cell types from 56 MS runs
across 8 acquisition batches. In addition to the 4 cell types (THP1,
U937, differentiated THP1 and U937), some batches also contain a mix
of THP1 and U937 cells. Samples were mostly run on the Orbitrap Fusion
Lumos Tribrid™ mass spectrometer, one batch was run on the Exploris
240 and another one on the timsTOF SCP. Raw MS data were preprocessed
using the \texttt{sage} software. An in-depth description of the dataset can
be found in the \emph{Material} section and \textbf{Table \ref{tab:datatab}}.

\begin{Shaded}
\begin{Highlighting}[]
\NormalTok{scp}
\end{Highlighting}
\end{Shaded}

\begin{verbatim}
## An instance of class QFeatures containing 56 assays:
##  [1] CBIO680_1: SingleCellExperiment with 39666 rows and 16 columns 
##  [2] CBIO680_3: SingleCellExperiment with 38191 rows and 16 columns 
##  [3] CBIO680_4: SingleCellExperiment with 36276 rows and 16 columns 
##  ...
##  [54] GIGA_1250: SingleCellExperiment with 531940 rows and 16 columns 
##  [55] GIGA_1251: SingleCellExperiment with 551645 rows and 16 columns 
##  [56] GIGA_1252: SingleCellExperiment with 363371 rows and 16 columns
\end{verbatim}

\hypertarget{missing-data}{%
\subsection{Missing data}\label{missing-data}}

The nature of mass spectrometry measurement and data processing leads
to ions not being detected or reported despite their presence at a
detectable level in the original samples. It is however not possible
to discriminate between values missing due to the absence of the
feature in the biological sample or for technical or analytical
reasons. The \texttt{sage} software, used for the processing of the raw mass
spectrometry data, reports those missing values as zeros. This leads
to an implicit imputation by 0 that should be avoided in MS-based
proteomics\textsuperscript{\protect\hyperlink{ref-Kong2022-wp}{24}}. The \texttt{zeroIsNA()} function replaces zeros
with \texttt{NAs} in every set.

\begin{Shaded}
\begin{Highlighting}[]
\NormalTok{scp }\OtherTok{\textless{}{-}} \FunctionTok{zeroIsNA}\NormalTok{(scp, }\AttributeTok{i =} \DecValTok{1}\SpecialCharTok{:}\FunctionTok{length}\NormalTok{(scp))}
\end{Highlighting}
\end{Shaded}

\hypertarget{quality-control}{%
\subsection{Quality control}\label{quality-control}}

In mass spectrometry-based proteomics, the raw data consist of spectra
with intensity peaks for a range of m/z values. Spectra are then
matched to their probabilistically most likely peptide sequence. Thus,
any spectrum that has been attributed to a peptide sequence is called
\emph{peptide to spectrum match} or PSM. This is the level in which our
processing starts before building our way to peptides and proteins. We
immediately start with a round of quality control (QC) to remove
poor-quality features and cells. Quality control is performed at the
PSM level to avoid the propagation of technical artefacts to the
downstream data.

\hypertarget{psms-filtering}{%
\subsubsection{PSMs filtering}\label{psms-filtering}}

A common step in SCP is to filter out low-confidence PSMs. Our
filtering relies on commonly used feature annotations provided by the
raw data processing software. In addition, we compute and use the
sample-to-carrier ratio (SCR)\textsuperscript{\protect\hyperlink{ref-specht_single-cell_2021}{16}}, a metric
specific to experiments using a carrier channel, when one is
available.

\hypertarget{filtering-based-on-features-annotations}{%
\paragraph{Filtering based on features annotations}\label{filtering-based-on-features-annotations}}

Each PSM set contains feature annotations that are stored in the
\texttt{rowData} slot of the set. The \texttt{QFeatures} package allows for a
streamlined filtering of the rows based on the information in the
\texttt{rowData}. This is done using the \texttt{filterFeatures()} function. Below,
we filter PSMs with rank 1 to only keep the sequences with the highest
score for each spectrum, and PSMs with a false discovery rate (FDR)
below 1\% as their identification is considered to be of sufficient
confidence. To estimate the false discovery rate, processing software
generate decoy peptides by reversing the protein sequence. \texttt{sage}
assigns reverse PSMs a value of -1 in the label column. Forward PSMs,
that have a label of 1, are retained.

\begin{Shaded}
\begin{Highlighting}[]
\NormalTok{scp }\OtherTok{\textless{}{-}} \FunctionTok{filterFeatures}\NormalTok{(scp,}
                      \SpecialCharTok{\textasciitilde{}}\NormalTok{ rank }\SpecialCharTok{==} \DecValTok{1} \SpecialCharTok{\&}
\NormalTok{                        peptide\_fdr }\SpecialCharTok{\textless{}} \FloatTok{0.01} \SpecialCharTok{\&}
\NormalTok{                        label }\SpecialCharTok{==} \DecValTok{1}\NormalTok{)}
\end{Highlighting}
\end{Shaded}

The sage software was configured to detect chimeric spectra. Under
this configuration, multiple identifications can be found for the same
spectrum. When performing isobaric multiplexing, it is not possible to
discriminate quantification originating from each peptide in the same
spectrum. We use the chimeric identification to filter out PSMs with
ambiguous quantification i.e.~PSMs that share a common spectrum
identifier.

\begin{itemize}
\tightlist
\item
  We create the spectrum-specific identifier \texttt{.KEY} by pasting the
  \texttt{file} and \texttt{scannr} variables.
\item
  We add a column called \texttt{chimeric} to the PSM annotations in the
  \texttt{rowData} slot.
\item
  We assign either \texttt{FALSE} if the scan identifier is unique or \texttt{TRUE}
  if it's not, to the \texttt{chimeric} column.
\item
  We store the updated annotations back into the \texttt{rowData} slot.
\end{itemize}

This process is looped for all sets of the \texttt{scp} object.

\begin{Shaded}
\begin{Highlighting}[]
\ControlFlowTok{for}\NormalTok{ (i }\ControlFlowTok{in} \FunctionTok{seq\_along}\NormalTok{(scp)) \{}
  \CommentTok{\# Extract rowData for each set}
\NormalTok{  rd }\OtherTok{\textless{}{-}} \FunctionTok{rowData}\NormalTok{(scp[[}\FunctionTok{names}\NormalTok{(scp)[i]]])}
  \CommentTok{\# Create unique spectrum identifier .KEY}
\NormalTok{  rd}\SpecialCharTok{$}\NormalTok{.KEY }\OtherTok{\textless{}{-}} \FunctionTok{paste}\NormalTok{(rd}\SpecialCharTok{$}\NormalTok{file, rd}\SpecialCharTok{$}\NormalTok{scannr)}
  \CommentTok{\# Create "chimeric" column, FALSE by default}
\NormalTok{  rd}\SpecialCharTok{$}\NormalTok{chimeric }\OtherTok{\textless{}{-}} \ConstantTok{FALSE}
  \CommentTok{\# Change "chimeric" to TRUE for duplicated keys}
\NormalTok{  rd}\SpecialCharTok{$}\NormalTok{chimeric[rd}\SpecialCharTok{$}\NormalTok{.KEY }\SpecialCharTok{\%in\%}\NormalTok{ rd}\SpecialCharTok{$}\NormalTok{.KEY[}\FunctionTok{duplicated}\NormalTok{(rd}\SpecialCharTok{$}\NormalTok{.KEY)]] }\OtherTok{\textless{}{-}} \ConstantTok{TRUE}
  \CommentTok{\# Store updated rowData}
  \FunctionTok{rowData}\NormalTok{(scp[[}\FunctionTok{names}\NormalTok{(scp)[i]]]) }\OtherTok{\textless{}{-}}\NormalTok{ rd}
\NormalTok{\}}
\end{Highlighting}
\end{Shaded}

\begin{Shaded}
\begin{Highlighting}[]
\FunctionTok{as.data.frame}\NormalTok{(}\FunctionTok{head}\NormalTok{(}\FunctionTok{rowData}\NormalTok{(scp[[}\DecValTok{1}\NormalTok{]]))[, }\FunctionTok{c}\NormalTok{(}\StringTok{".KEY"}\NormalTok{, }\StringTok{"peptide"}\NormalTok{, }\StringTok{"chimeric"}\NormalTok{)])}
\end{Highlighting}
\end{Shaded}

\begin{verbatim}
##                                 .KEY       peptide chimeric
## PSM3039205 CBIO680_1.mzML scan=21123        LSGLPK    FALSE
## PSM3039231 CBIO680_1.mzML scan=21158 QADLYISEGLHPR    FALSE
## PSM3039266 CBIO680_1.mzML scan=21194   AVFPSIVGRPR     TRUE
## PSM3039267 CBIO680_1.mzML scan=21194      EAILAIHK     TRUE
## PSM3039277 CBIO680_1.mzML scan=21206        LLVGNK    FALSE
## PSM3039313 CBIO680_1.mzML scan=21241      FFPASADR    FALSE
\end{verbatim}

Features highlighted as chimeric are removed using \texttt{filterFeatures()}.
In this dataset, a median of 177 chimeric PSMs were filtered out,
representing 5.2\% of total number of PSMs at this stage.

\begin{Shaded}
\begin{Highlighting}[]
\NormalTok{scp }\OtherTok{\textless{}{-}} \FunctionTok{filterFeatures}\NormalTok{(scp,}
                      \SpecialCharTok{\textasciitilde{}} \SpecialCharTok{!}\NormalTok{chimeric)}
\end{Highlighting}
\end{Shaded}

\hypertarget{filtering-based-on-scp-metrics}{%
\paragraph{Filtering based on SCP metrics}\label{filtering-based-on-scp-metrics}}

The next filter is based on the sample-to-carrier ratio (SCR). SCR is
the ratio of the reporter ion intensity of a single cell and the
reporter ion intensity of the carrier channel (here 50 cells) from the
same batch. We expect the carrier intensities to be about 50x higher
than the single-cell intensities, hence we expect the SCRs to be on average 1/50. Note that
the average ratio of 1/50 is theoretical; in practice, noise and ratio
compression are likely to shift ratios towards higher values.

The SCRs can be computed using the \texttt{computeSCR()} function from
\texttt{scp}. The function must be told which channels are the cells and
which channel is the carrier. This information is available in the
\texttt{cell\_type} variable in the object's cell annotations

\begin{Shaded}
\begin{Highlighting}[]
\FunctionTok{table}\NormalTok{(scp}\SpecialCharTok{$}\NormalTok{cell\_type)}
\end{Highlighting}
\end{Shaded}

\begin{verbatim}
## 
##    blank  carrier    empty      mix     THP1 THP1_dif     U937 U937_dif 
##      103       56      125       56      188       90      188       90
\end{verbatim}

We consider the quantification of THP1, THP1 differentiated, U937,
U937 differentiated and mix as single cells. Blank and empty samples
are not considered as they contain no cells and should not comply with
an SCR of around 1/50. For each PSM, the function averages the SCRs of
all cells. Finally, the average SCRs for each PSM are stored with the
feature annotations in the \texttt{rowData} slot.

\begin{Shaded}
\begin{Highlighting}[]
\NormalTok{scp }\OtherTok{\textless{}{-}} \FunctionTok{computeSCR}\NormalTok{(scp,}
                  \AttributeTok{i =} \DecValTok{1}\SpecialCharTok{:}\FunctionTok{length}\NormalTok{(scp),}
                  \AttributeTok{colvar =} \StringTok{"cell\_type"}\NormalTok{,}
                  \AttributeTok{carrierPattern =} \StringTok{"carrier"}\NormalTok{,}
                  \AttributeTok{samplePattern =} \StringTok{"THP1|THP1\_dif|U937|U937\_dif|mix"}\NormalTok{,}
                  \AttributeTok{rowDataName =} \StringTok{"MeanSCR"}\NormalTok{)}
\end{Highlighting}
\end{Shaded}

\begin{figure}
\centering
\includegraphics{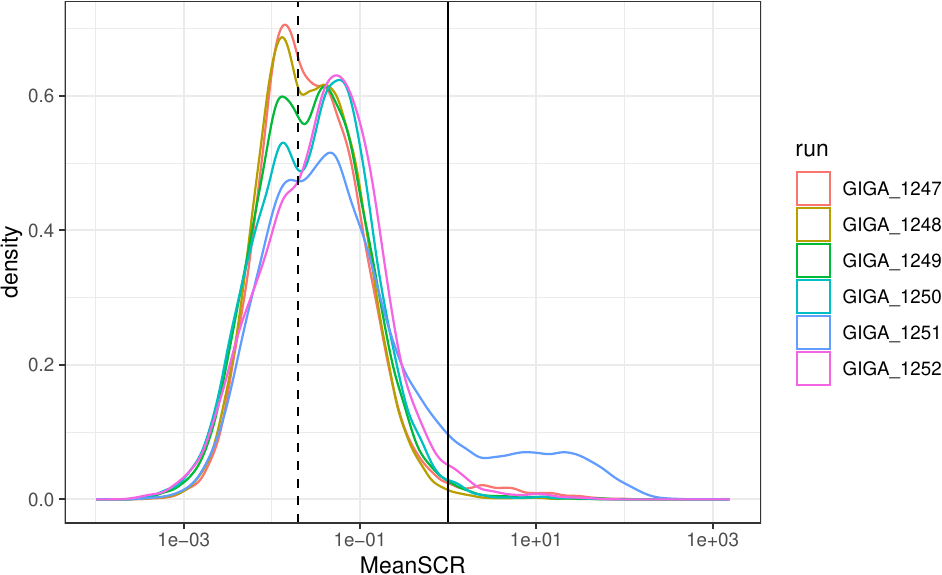}
\caption{\label{fig:plotscr}SCR distributions from the 6 runs of the GIGA batch. SCR distributions are centred around the expected 1/50 ratio (dashed vertical line). Threshold is set at 1 (solid vertical line) to remove PSMs with unexpectedly high SCRs.}
\end{figure}

Before applying the filter, the distribution of the average SCRs is
plotted (\textbf{Figure \ref{fig:plotscr}}). The feature annotations from
several sets are collected in a single table using the
\texttt{rbindRowData()} function from \texttt{QFeatures}. The plot below focuses on
the acquisition from the GIGA batch.

\begin{Shaded}
\begin{Highlighting}[]
\FunctionTok{rbindRowData}\NormalTok{(scp, }\AttributeTok{i =} \DecValTok{1}\SpecialCharTok{:}\FunctionTok{length}\NormalTok{(scp))  }\SpecialCharTok{|\textgreater{}}
  \FunctionTok{data.frame}\NormalTok{() }\SpecialCharTok{|\textgreater{}}
  \FunctionTok{filter}\NormalTok{(batch }\SpecialCharTok{==} \StringTok{"GIGA"}\NormalTok{) }\SpecialCharTok{|\textgreater{}}
  \FunctionTok{ggplot}\NormalTok{(}\FunctionTok{aes}\NormalTok{(}\AttributeTok{x =}\NormalTok{ MeanSCR, }\AttributeTok{color =}\NormalTok{ run)) }\SpecialCharTok{+}
  \FunctionTok{geom\_density}\NormalTok{() }\SpecialCharTok{+}
  \FunctionTok{geom\_vline}\NormalTok{(}\AttributeTok{xintercept =} \FloatTok{0.02}\NormalTok{,}
             \AttributeTok{lty =} \DecValTok{2}\NormalTok{) }\SpecialCharTok{+}
  \FunctionTok{geom\_vline}\NormalTok{(}\AttributeTok{xintercept =} \DecValTok{1}\NormalTok{,}
             \AttributeTok{lty =} \DecValTok{1}\NormalTok{)}\SpecialCharTok{+}
  \FunctionTok{scale\_x\_log10}\NormalTok{()}
\end{Highlighting}
\end{Shaded}

For this batch, the SCRs are mostly centred on the expected 1/50
ratio. Note that this is not always the case, especially in case of
important losses during sample preparation. In run ``GIGA\_1251''
specifically, a few PSMs stand out of the distribution and have a much
higher signal than expected, indicating that caution is required
regarding the quantification of those PSMs. They are filtered out with
a threshold set at 1 (solid vertical line). This is again easily
performed using the \texttt{filterFeatures()} function. The aim of this
filtering step is to eliminate outlier PSMs rather than focusing on
the SCR values themselves. We found that a stricter SCR threshold
reduced the number of features without improving data quality.

\begin{Shaded}
\begin{Highlighting}[]
\NormalTok{scp }\OtherTok{\textless{}{-}} \FunctionTok{filterFeatures}\NormalTok{(scp,}
                      \SpecialCharTok{\textasciitilde{}} \SpecialCharTok{!}\FunctionTok{is.na}\NormalTok{(MeanSCR) }\SpecialCharTok{\&}
\NormalTok{                          MeanSCR }\SpecialCharTok{\textless{}} \DecValTok{1}\NormalTok{)}
\end{Highlighting}
\end{Shaded}

Note that PSMs only found in the carrier channel have missing values
for their SCR. They are also removed during this step. Filtering
based on FDR removed a median of 240 additional PSMs, representing
7.7\% of total PSMs at this stage.

\hypertarget{cell-filtering}{%
\subsubsection{Cell filtering}\label{cell-filtering}}

After removing low-quality features in the previous section, we now
perform a quality control for the cells. We remove irrelevant samples
and apply a filter based on 3 metrics: number of peptides, median
reporter intensity (RI) and median coefficient of variation (CV).

\hypertarget{removing-irrelevant-samples}{%
\paragraph{Removing irrelevant samples}\label{removing-irrelevant-samples}}

From this point on, carrier and empty channels are no longer useful
and can be discarded. Again, this step is streamlined thanks to the
\texttt{subsetByColData()} function, which discards cells based on the cell
annotations.

\begin{Shaded}
\begin{Highlighting}[]
\FunctionTok{table}\NormalTok{(scp}\SpecialCharTok{$}\NormalTok{cell\_type)}
\end{Highlighting}
\end{Shaded}

\begin{verbatim}
## 
##    blank  carrier    empty      mix     THP1 THP1_dif     U937 U937_dif 
##      103       56      125       56      188       90      188       90
\end{verbatim}

\begin{Shaded}
\begin{Highlighting}[]
\NormalTok{scp }\OtherTok{\textless{}{-}} \FunctionTok{subsetByColData}\NormalTok{(scp, }\SpecialCharTok{!}\NormalTok{scp}\SpecialCharTok{$}\NormalTok{cell\_type }\SpecialCharTok{\%in\%} \FunctionTok{c}\NormalTok{(}\StringTok{"carrier"}\NormalTok{, }\StringTok{"empty"}\NormalTok{))}

\FunctionTok{table}\NormalTok{(scp}\SpecialCharTok{$}\NormalTok{cell\_type)}
\end{Highlighting}
\end{Shaded}

\begin{verbatim}
## 
##    blank      mix     THP1 THP1_dif     U937 U937_dif 
##      103       56      188       90      188       90
\end{verbatim}

This way, only single-cell samples and blanks, used for quality
control, remain.

Note that samples are grouped by batches during the following
filtering (see \emph{Note 1}). The filtering are only illustrated for
batches CBIO715 and CBIO681. However, we applied the filtering
similarly to all batches.

\hypertarget{filtering-based-on-median-reporter-intensity}{%
\paragraph{Filtering based on median reporter intensity}\label{filtering-based-on-median-reporter-intensity}}

The median reporter ion intensity (RI) is computed for each cell
separately using the \texttt{colMedians()} function. This information is
stored with the cell annotations in the \texttt{colData} slot so that a
filter can be applied based on this metric in subsequent steps.

\begin{Shaded}
\begin{Highlighting}[]
\ControlFlowTok{for}\NormalTok{ (i }\ControlFlowTok{in} \FunctionTok{names}\NormalTok{(scp)) \{}
  \CommentTok{\# Extract log assay}
\NormalTok{  logAssay }\OtherTok{\textless{}{-}} \FunctionTok{log}\NormalTok{(}\FunctionTok{assay}\NormalTok{(scp[[i]]))}
  \CommentTok{\# Compute median RI by cell}
\NormalTok{  meds }\OtherTok{\textless{}{-}} \FunctionTok{colMedians}\NormalTok{(logAssay, }\AttributeTok{na.rm =} \ConstantTok{TRUE}\NormalTok{, }\AttributeTok{useNames =} \ConstantTok{TRUE}\NormalTok{)}
  \CommentTok{\# Store median RI in colData.}
  \FunctionTok{colData}\NormalTok{(scp)[}\FunctionTok{names}\NormalTok{(meds), }\StringTok{"log\_medianRI"}\NormalTok{] }\OtherTok{\textless{}{-}}\NormalTok{ meds}
\NormalTok{\}}
\end{Highlighting}
\end{Shaded}

\begin{figure}
\centering
\includegraphics{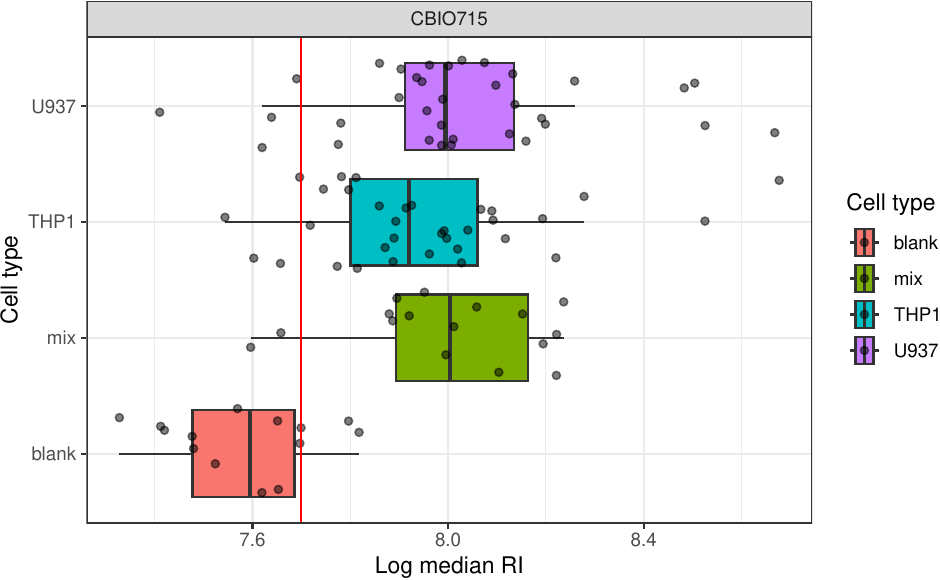}
\caption{\label{fig:plotRI}Distribution of log median RI per cell types. U937, THP1, mix, and blank log median RI distributions. Blanks distribution is used to estimate the background noise and to set the filtering threshold (red vertical line).}
\end{figure}

To help us decide which threshold to use, the distributions of the
median RI are plotted for each cell type (\textbf{Figure
\ref{fig:plotRI}}). The filter is shown for batches CBIO715, but we
applied a similar filter to all batches individually (see \emph{Note 1}).
The negative control samples (blanks) do not contain any cells and are
therefore used to assess the amount of background signal. Here, the
signal measured in most single cells is above the background signal,
which is however not always the case (see \emph{Note 2}). Based on the
blank distribution, a threshold is set and single cells with the
median RI below the threshold will be removed.

\begin{Shaded}
\begin{Highlighting}[]
\FunctionTok{colData}\NormalTok{(scp) }\SpecialCharTok{|\textgreater{}}
  \FunctionTok{data.frame}\NormalTok{() }\SpecialCharTok{|\textgreater{}}
  \FunctionTok{filter}\NormalTok{(batch }\SpecialCharTok{==} \StringTok{"CBIO715"}\NormalTok{) }\SpecialCharTok{|\textgreater{}}
  \FunctionTok{ggplot}\NormalTok{() }\SpecialCharTok{+}
  \FunctionTok{aes}\NormalTok{(}\AttributeTok{x =}\NormalTok{ log\_medianRI,}
      \AttributeTok{y =}\NormalTok{ cell\_type,}
      \AttributeTok{fill =}\NormalTok{ cell\_type) }\SpecialCharTok{+}
  \FunctionTok{geom\_boxplot}\NormalTok{(}\AttributeTok{outlier.shape =} \ConstantTok{NA}\NormalTok{) }\SpecialCharTok{+}
  \FunctionTok{geom\_jitter}\NormalTok{(}\AttributeTok{alpha =} \FloatTok{0.5}\NormalTok{) }\SpecialCharTok{+}
  \FunctionTok{facet\_wrap}\NormalTok{(}\SpecialCharTok{\textasciitilde{}}\NormalTok{ batch) }\SpecialCharTok{+}
  \FunctionTok{labs}\NormalTok{(}\AttributeTok{fill =} \StringTok{"Cell type"}\NormalTok{,}
       \AttributeTok{y =} \StringTok{"Cell type"}\NormalTok{,}
       \AttributeTok{x =} \StringTok{"Log median RI"}\NormalTok{) }\SpecialCharTok{+}
  \FunctionTok{geom\_vline}\NormalTok{(}\AttributeTok{xintercept =} \FloatTok{7.7}\NormalTok{,}
             \AttributeTok{color =} \StringTok{"red"}\NormalTok{)}
\end{Highlighting}
\end{Shaded}

\hypertarget{filtering-based-on-the-median-coefficient-of-variation}{%
\paragraph{Filtering based on the median coefficient of variation}\label{filtering-based-on-the-median-coefficient-of-variation}}

The median coefficient of variation (CV) measures the consistency of
quantification for a set of PSMs belonging to a protein within a
cell. The coefficient of variation is defined by the ratio between the
standard deviation and the mean. The \texttt{computeMedianCV()} function from
the \texttt{scp} package computes the CV for each protein in each cell. The
CVs are then summarised for each cell using the median. PSM to protein
assignment is defined by the \texttt{proteins} variable in the features
annotations (\texttt{rowData}) through the \texttt{groupBy} argument. CVs are only
computed if there are at least 3 PSMs per protein (\texttt{nobs}
argument). Since multiple PSMs of different peptides are used to
calculate the CV of a protein, each row in each \texttt{assay} needs to be
normalised using the method provided by the \texttt{norm} argument. The
computed median CVs are automatically stored with the cell annotations
in the \texttt{colData} slot under the name that is supplied for
\texttt{colDataName}, here \texttt{medianCV}.

\begin{Shaded}
\begin{Highlighting}[]
\NormalTok{scp }\OtherTok{\textless{}{-}} \FunctionTok{medianCVperCell}\NormalTok{(scp,}
                       \AttributeTok{i =} \DecValTok{1}\SpecialCharTok{:}\FunctionTok{length}\NormalTok{(scp),}
                       \AttributeTok{groupBy =} \StringTok{"proteins"}\NormalTok{,}
                       \AttributeTok{nobs =} \DecValTok{3}\NormalTok{,}
                       \AttributeTok{norm =} \StringTok{"div.median"}\NormalTok{,}
                       \AttributeTok{colDataName =} \StringTok{"medianCV"}\NormalTok{)}
\end{Highlighting}
\end{Shaded}

\begin{figure}
\centering
\includegraphics{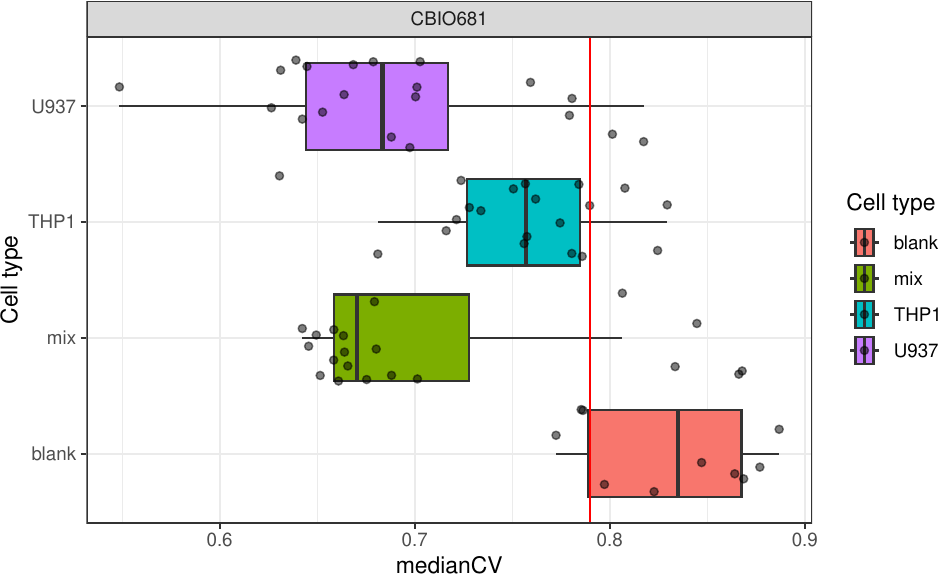}
\caption{\label{fig:plotCV}Distribution of median CV per cell types. U937, THP1, mix, and blank log median CV distributions. Blanks distribution is used to estimate CVs in background noise and to set the filtering threshold (red vertical line).}
\end{figure}

Similarly to median RI, median CV distributions are plotted for each
cell type (\textbf{Figure \ref{fig:plotCV}}). Again, the filter is shown
for batches CBIO681, but we applied a similar filter to all batches
individually (see \emph{Note 1}). The main interest of computing the
median CV per cell is to remove cells with unreliable
quantification. In reliable single-cell samples, we expect only a
slight variation in quantification for PSMs belonging to the same
protein. However, the negative control should only contain noise and
no consistency in PSMs quantification is expected. Therefore, negative
control samples are used to estimate an empirical null distribution of
the CV. This distribution helps defining a threshold that filters out
single cells containing noisy quantification.

\begin{Shaded}
\begin{Highlighting}[]
\FunctionTok{colData}\NormalTok{(scp) }\SpecialCharTok{|\textgreater{}}
  \FunctionTok{data.frame}\NormalTok{() }\SpecialCharTok{|\textgreater{}}
  \FunctionTok{filter}\NormalTok{(batch }\SpecialCharTok{==} \StringTok{"CBIO681"}\NormalTok{) }\SpecialCharTok{|\textgreater{}}
  \FunctionTok{ggplot}\NormalTok{() }\SpecialCharTok{+}
  \FunctionTok{aes}\NormalTok{(}\AttributeTok{x =}\NormalTok{ medianCV,}
      \AttributeTok{y =}\NormalTok{ cell\_type,}
      \AttributeTok{fill =}\NormalTok{ cell\_type) }\SpecialCharTok{+}
  \FunctionTok{geom\_boxplot}\NormalTok{(}\AttributeTok{outlier.shape =} \ConstantTok{NA}\NormalTok{) }\SpecialCharTok{+}
  \FunctionTok{geom\_jitter}\NormalTok{(}\AttributeTok{alpha =} \FloatTok{0.5}\NormalTok{)}\SpecialCharTok{+}
  \FunctionTok{facet\_wrap}\NormalTok{(}\SpecialCharTok{\textasciitilde{}}\NormalTok{batch)}\SpecialCharTok{+}
  \FunctionTok{labs}\NormalTok{(}\AttributeTok{fill =} \StringTok{"Cell type"}\NormalTok{,}
       \AttributeTok{y =} \StringTok{"Cell type"}\NormalTok{) }\SpecialCharTok{+}
  \FunctionTok{geom\_vline}\NormalTok{(}\AttributeTok{xintercept =} \FloatTok{0.79}\NormalTok{,}
             \AttributeTok{color =} \StringTok{"red"}\NormalTok{)}
\end{Highlighting}
\end{Shaded}

\hypertarget{filtering-based-on-peptide-numbers}{%
\paragraph{Filtering based on peptide numbers}\label{filtering-based-on-peptide-numbers}}

We count the number of peptides in each cell using the
\texttt{countUniqueFeatures()} function. Similarly to \texttt{medianCVperCell()}, we
use the \texttt{groupBy} argument to indicate which features annotations
field from the \texttt{rowData} slot to use to group the PSMs into
peptides. Peptide numbers are automatically stored with the cell
annotations in the \texttt{colData} slot under the \texttt{count} name. We keep
cells with more than 1250 peptides since, knowing the performance of
our methods, identifying less than 1250 peptides should only happen in
poor-quality cell samples. In addition, having a low number of
peptides only gives us limited insights into the proteome.

\begin{Shaded}
\begin{Highlighting}[]
\NormalTok{scp }\OtherTok{\textless{}{-}} \FunctionTok{countUniqueFeatures}\NormalTok{(scp,}
                           \AttributeTok{i =} \DecValTok{1}\SpecialCharTok{:}\FunctionTok{length}\NormalTok{(scp),}
                           \AttributeTok{groupBy =} \StringTok{"peptide"}\NormalTok{,}
                           \AttributeTok{colDataName =} \StringTok{"count"}\NormalTok{)}
\FunctionTok{head}\NormalTok{(scp}\SpecialCharTok{$}\NormalTok{count)}
\end{Highlighting}
\end{Shaded}

\begin{verbatim}
## [1] 1168 1239 1325 1384 1347 1240
\end{verbatim}

\hypertarget{quality-control-overview}{%
\subsubsection{Quality control overview}\label{quality-control-overview}}

\begin{figure}
\centering
\includegraphics{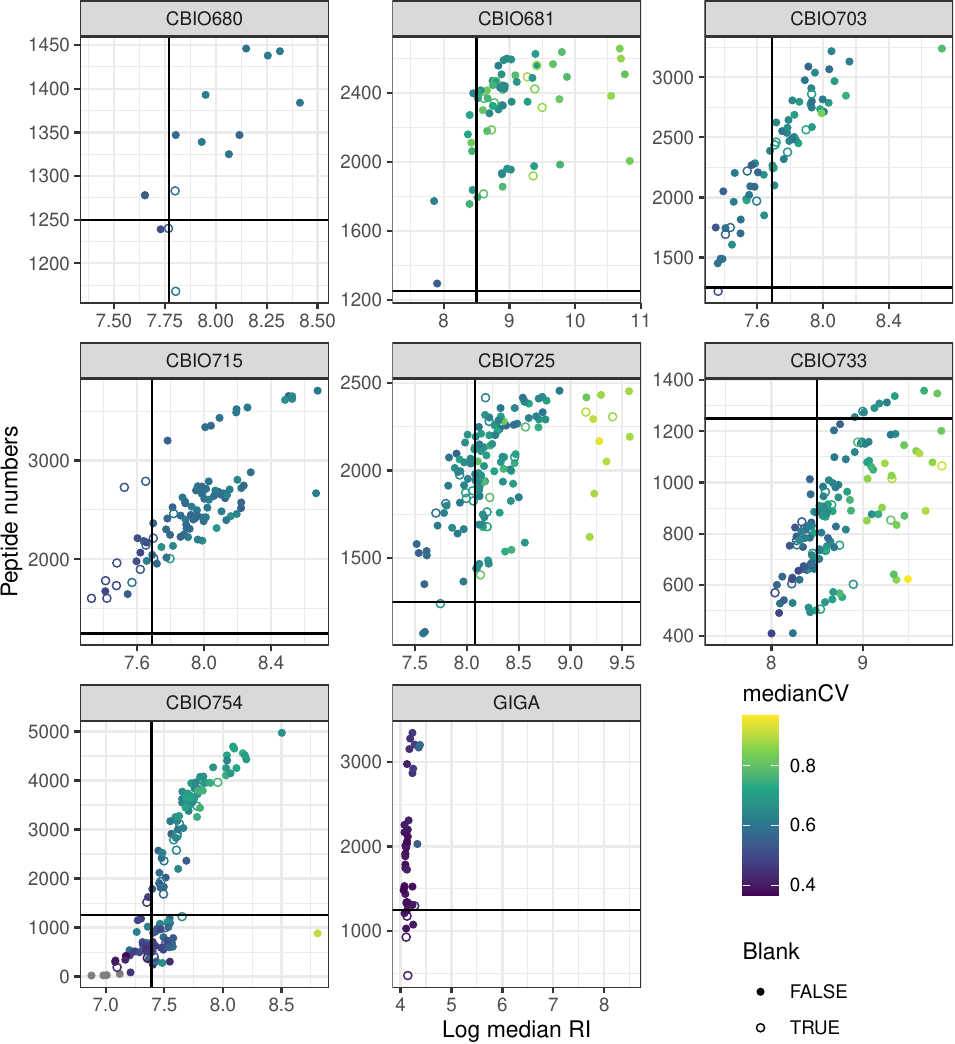}
\caption{\label{fig:QC}Quality Control overview. Cells are plotted based on their log median RI and Peptide numbers. Cells are colored based on their median CV. Threshold for log median RI and peptide numbers filtering are shown by vertical and horizontal lines, respectively.}
\end{figure}

To get a global overview of the quality control (QC), we plot the 3
metrics with their corresponding thresholds (\textbf{Figure
\ref{fig:QC}}). Cells in the upper right corner, and with a low
enough CV will be kept.

\begin{Shaded}
\begin{Highlighting}[]
\NormalTok{scp }\SpecialCharTok{|\textgreater{}}
  \FunctionTok{colData}\NormalTok{() }\SpecialCharTok{|\textgreater{}}
  \FunctionTok{as.data.frame}\NormalTok{() }\SpecialCharTok{|\textgreater{}}
  \FunctionTok{ggplot}\NormalTok{(}\FunctionTok{aes}\NormalTok{(}\AttributeTok{x =}\NormalTok{ log\_medianRI, }\AttributeTok{y =}\NormalTok{ count,}
             \AttributeTok{color =}\NormalTok{ medianCV, }\AttributeTok{shape =}\NormalTok{ cell\_type }\SpecialCharTok{==} \StringTok{"blank"}\NormalTok{)) }\SpecialCharTok{+}
  \FunctionTok{geom\_point}\NormalTok{() }\SpecialCharTok{+}
  \FunctionTok{scale\_color\_viridis\_c}\NormalTok{() }\SpecialCharTok{+}
  \FunctionTok{facet\_wrap}\NormalTok{(}\SpecialCharTok{\textasciitilde{}}\NormalTok{ batch, }\AttributeTok{scales =} \StringTok{"free"}\NormalTok{) }\SpecialCharTok{+}
  \FunctionTok{geom\_vline}\NormalTok{(}\AttributeTok{xintercept =} \FunctionTok{c}\NormalTok{(}\FloatTok{7.77}\NormalTok{, }\FloatTok{8.5}\NormalTok{, }\FloatTok{7.69}\NormalTok{, }\FloatTok{7.69}\NormalTok{, }\FloatTok{8.08}\NormalTok{, }\FloatTok{8.5}\NormalTok{, }\FloatTok{7.39}\NormalTok{, }\ConstantTok{NA}\NormalTok{),}
             \AttributeTok{lty =} \FunctionTok{c}\NormalTok{(}\FunctionTok{diag}\NormalTok{(}\DecValTok{1}\NormalTok{, }\DecValTok{8}\NormalTok{, }\DecValTok{8}\NormalTok{))) }\SpecialCharTok{+}
  \FunctionTok{geom\_hline}\NormalTok{(}\AttributeTok{yintercept =} \DecValTok{1250}\NormalTok{) }\SpecialCharTok{+}
  \FunctionTok{scale\_shape\_manual}\NormalTok{(}\AttributeTok{values =} \FunctionTok{c}\NormalTok{(}\DecValTok{16}\NormalTok{, }\DecValTok{21}\NormalTok{)) }\SpecialCharTok{+}
  \FunctionTok{labs}\NormalTok{(}\AttributeTok{shape =} \StringTok{"Blank"}\NormalTok{,}
       \AttributeTok{y =} \StringTok{"Peptide numbers"}\NormalTok{,}
       \AttributeTok{x =} \StringTok{"Log median RI"}\NormalTok{,}
       \AttributeTok{fill =} \StringTok{"median CV"}\NormalTok{) }\SpecialCharTok{+}
  \FunctionTok{theme}\NormalTok{(}\AttributeTok{legend.position =} \FunctionTok{c}\NormalTok{(}\FloatTok{0.82}\NormalTok{, }\FloatTok{0.13}\NormalTok{))}
\end{Highlighting}
\end{Shaded}

Batch \texttt{CBIO733} shows poor quality with only a few cells exceeding
1250 peptides. This might be due to the poor calibration of the mass
spectrometer since it was the first time any single-cell samples were
acquired on this instrument. The complete batch will be removed in the
next step. Batch \texttt{GIGA} was run on a different type of mass
spectrometer (see \textbf{Table \ref{tab:datatab}}), explaining why the
range of intensities is different. In addition, all cells are
concentrated around the same acceptable median RI, so median RI will
not be used to filter cells in this batch.

Once thresholds have been defined, cells that pass all quality
controls are retained. This is done by extracting the relevant metrics
from the cell annotations. The cells that pass the filters are kept
using \texttt{subsetByColData()} once more.

\begin{Shaded}
\begin{Highlighting}[]
\NormalTok{filter\_samples }\OtherTok{\textless{}{-}}
\NormalTok{  (scp}\SpecialCharTok{$}\NormalTok{batch }\SpecialCharTok{==} \StringTok{"CBIO680"} \SpecialCharTok{\&}\NormalTok{ scp}\SpecialCharTok{$}\NormalTok{log\_medianRI }\SpecialCharTok{\textgreater{}} \FloatTok{7.77} \SpecialCharTok{\&}
\NormalTok{     scp}\SpecialCharTok{$}\NormalTok{count }\SpecialCharTok{\textgreater{}} \DecValTok{1250} \SpecialCharTok{\&}\NormalTok{ scp}\SpecialCharTok{$}\NormalTok{medianCV }\SpecialCharTok{\textless{}} \FloatTok{0.615}\NormalTok{) }\SpecialCharTok{|}
\NormalTok{  (scp}\SpecialCharTok{$}\NormalTok{batch }\SpecialCharTok{==} \StringTok{"CBIO681"} \SpecialCharTok{\&}\NormalTok{ scp}\SpecialCharTok{$}\NormalTok{log\_medianRI }\SpecialCharTok{\textgreater{}} \FloatTok{8.5} \SpecialCharTok{\&}
\NormalTok{     scp}\SpecialCharTok{$}\NormalTok{count }\SpecialCharTok{\textgreater{}} \DecValTok{1250} \SpecialCharTok{\&}\NormalTok{ scp}\SpecialCharTok{$}\NormalTok{medianCV }\SpecialCharTok{\textless{}} \FloatTok{0.79}\NormalTok{) }\SpecialCharTok{|}
\NormalTok{  (scp}\SpecialCharTok{$}\NormalTok{batch }\SpecialCharTok{==} \StringTok{"CBIO703"} \SpecialCharTok{\&}\NormalTok{ scp}\SpecialCharTok{$}\NormalTok{log\_medianRI }\SpecialCharTok{\textgreater{}} \FloatTok{7.69} \SpecialCharTok{\&}
\NormalTok{     scp}\SpecialCharTok{$}\NormalTok{count }\SpecialCharTok{\textgreater{}} \DecValTok{1250} \SpecialCharTok{\&}\NormalTok{ scp}\SpecialCharTok{$}\NormalTok{medianCV }\SpecialCharTok{\textless{}} \FloatTok{0.68}\NormalTok{) }\SpecialCharTok{|}
\NormalTok{  (scp}\SpecialCharTok{$}\NormalTok{batch }\SpecialCharTok{==} \StringTok{"CBIO715"} \SpecialCharTok{\&}\NormalTok{ scp}\SpecialCharTok{$}\NormalTok{log\_medianRI }\SpecialCharTok{\textgreater{}} \FloatTok{7.69} \SpecialCharTok{\&}
\NormalTok{     scp}\SpecialCharTok{$}\NormalTok{count }\SpecialCharTok{\textgreater{}} \DecValTok{1250} \SpecialCharTok{\&}\NormalTok{ scp}\SpecialCharTok{$}\NormalTok{medianCV }\SpecialCharTok{\textless{}} \FloatTok{0.62}\NormalTok{) }\SpecialCharTok{|}
\NormalTok{  (scp}\SpecialCharTok{$}\NormalTok{batch }\SpecialCharTok{==} \StringTok{"CBIO725"} \SpecialCharTok{\&}\NormalTok{ scp}\SpecialCharTok{$}\NormalTok{log\_medianRI }\SpecialCharTok{\textgreater{}} \FloatTok{8.08} \SpecialCharTok{\&}
\NormalTok{     scp}\SpecialCharTok{$}\NormalTok{count }\SpecialCharTok{\textgreater{}} \DecValTok{1250} \SpecialCharTok{\&}\NormalTok{ scp}\SpecialCharTok{$}\NormalTok{medianCV }\SpecialCharTok{\textless{}} \FloatTok{0.73}\NormalTok{) }\SpecialCharTok{|}
\NormalTok{  (scp}\SpecialCharTok{$}\NormalTok{batch }\SpecialCharTok{==} \StringTok{"CBIO754"} \SpecialCharTok{\&}\NormalTok{ scp}\SpecialCharTok{$}\NormalTok{log\_medianRI }\SpecialCharTok{\textgreater{}} \FloatTok{7.39} \SpecialCharTok{\&}
\NormalTok{     scp}\SpecialCharTok{$}\NormalTok{count }\SpecialCharTok{\textgreater{}} \DecValTok{1250} \SpecialCharTok{\&}\NormalTok{ scp}\SpecialCharTok{$}\NormalTok{medianCV }\SpecialCharTok{\textless{}} \FloatTok{0.67}\NormalTok{) }\SpecialCharTok{|}
\NormalTok{  (scp}\SpecialCharTok{$}\NormalTok{batch }\SpecialCharTok{==} \StringTok{"GIGA"} \SpecialCharTok{\&}
\NormalTok{     scp}\SpecialCharTok{$}\NormalTok{count }\SpecialCharTok{\textgreater{}} \DecValTok{1250} \SpecialCharTok{\&}\NormalTok{ scp}\SpecialCharTok{$}\NormalTok{medianCV }\SpecialCharTok{\textless{}} \FloatTok{0.455}\NormalTok{)}

\NormalTok{scp }\OtherTok{\textless{}{-}} \FunctionTok{subsetByColData}\NormalTok{(scp, filter\_samples) }\SpecialCharTok{|\textgreater{}}
    \FunctionTok{dropEmptyAssays}\NormalTok{()}
\NormalTok{scp}
\end{Highlighting}
\end{Shaded}

\begin{verbatim}
## An instance of class QFeatures containing 42 assays:
##  [1] CBIO680_1: SingleCellExperiment with 1862 rows and 3 columns 
##  [2] CBIO680_3: SingleCellExperiment with 1941 rows and 1 columns 
##  [3] CBIO680_4: SingleCellExperiment with 1948 rows and 4 columns 
##  ...
##  [40] GIGA_1250: SingleCellExperiment with 31330 rows and 6 columns 
##  [41] GIGA_1251: SingleCellExperiment with 23046 rows and 4 columns 
##  [42] GIGA_1252: SingleCellExperiment with 16299 rows and 2 columns
\end{verbatim}

Notice how the number of columns is reduced as a result of
filtering. We also have fewer sets since empty sets have been removed.

Cell populations showing both high median RIs and CVs (mainly present
in batches CBIO725 and CBIO733) are suspected of being samples that
have undergone extensive contamination during preparation.
Contamination artificially boosts the signal and disrupts
quantification.

After filtering, remaining blanks are not useful anymore and can also
be discarded.

\begin{Shaded}
\begin{Highlighting}[]
\NormalTok{scp }\OtherTok{\textless{}{-}} \FunctionTok{subsetByColData}\NormalTok{(scp, scp}\SpecialCharTok{$}\NormalTok{cell\_type }\SpecialCharTok{!=} \StringTok{"blank"}\NormalTok{)}
\end{Highlighting}
\end{Shaded}

\hypertarget{peptide-data-assembling}{%
\subsubsection{Peptide data assembling}\label{peptide-data-assembling}}

Now that low-quality PSMs have been removed, the remaining PSMs can be
aggregated into peptides. This is performed using the
\texttt{aggregateFeatures()} function. For each set, the function aggregates
PSMs matched to the same sequence into a peptide. We provide the
feature variable to use for aggregation, i.e.~the peptide sequences,
using the \texttt{fcol} argument. We also need to supply an aggregating
function that defines how to compute the peptide-level quantitative
data from the PSM data with the \texttt{fun} argument. Here we use the
median.

\begin{Shaded}
\begin{Highlighting}[]
\NormalTok{scp }\OtherTok{\textless{}{-}} \FunctionTok{aggregateFeatures}\NormalTok{(scp,}
                         \AttributeTok{i =} \DecValTok{1}\SpecialCharTok{:}\FunctionTok{length}\NormalTok{(scp),}
                         \AttributeTok{fcol =} \StringTok{"peptide"}\NormalTok{,}
                         \AttributeTok{name =} \FunctionTok{paste0}\NormalTok{(}\StringTok{"peptide\_"}\NormalTok{, }\FunctionTok{names}\NormalTok{(scp)),}
                         \AttributeTok{fun =}\NormalTok{ colMedians, }\AttributeTok{na.rm =} \ConstantTok{TRUE}\NormalTok{)}
\end{Highlighting}
\end{Shaded}

\begin{figure}
\centering
\includegraphics{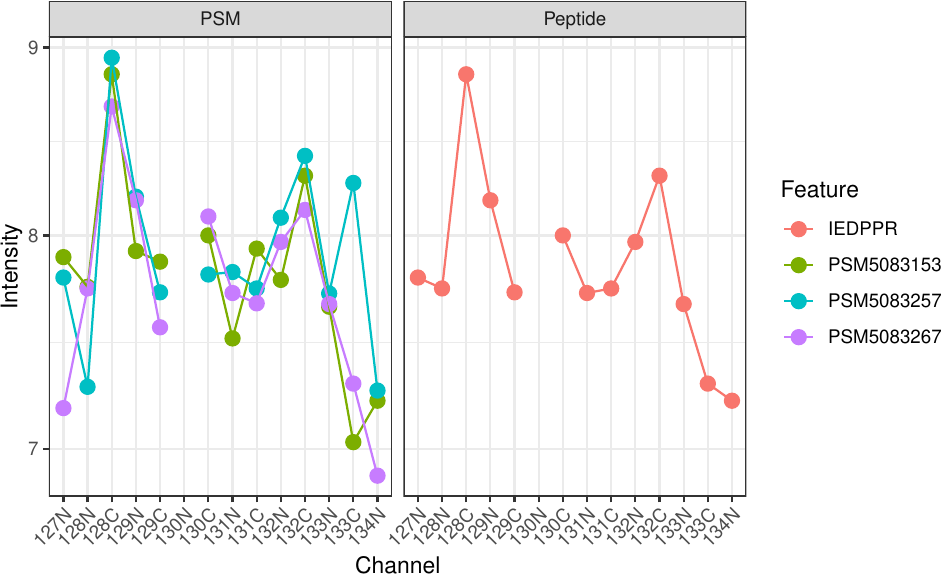}
\caption{\label{fig:aggregationExample}Example of PSM to peptide aggregation. PSM intensities for each channel are plotted in the left pannel. Intensities of the peptide resulting from the aggregation of the PSMs are plotted in the right pannel.}
\end{figure}

The \texttt{aggregateFeatures()} function creates a new set for each
aggregated set. The aggregated sets are named using the original names
and appending ``peptide\_'' at the start. \textbf{Figure
\ref{fig:aggregationExample}} illustrates the aggregation for three
PSMs that were matched to the same peptide sequence.

All sets belonging to the same batch are combined into a single
set. The combined sets will have as many columns as the sum of the
columns in the individual sets to join. All features found in at least
one sample will be part of the combined sets, which means that missing
values are added in columns (cells) from sets where the features were
absent. The joining is done using the \texttt{joinAssays()} function from the
\texttt{QFeatures} package. A loop is created to sequentially join sets from
the same batch. We retrieve the indexes for these sets by pasting
together ``peptide\_'' and the name of the batches and finding the
position of matches in all set names using \texttt{grep()}. The names of the
newly joined sets are created by pasting ``peptides\_'' to the name of
the batches.

\begin{Shaded}
\begin{Highlighting}[]
\NormalTok{batches }\OtherTok{\textless{}{-}} \FunctionTok{c}\NormalTok{(}\StringTok{"CBIO680"}\NormalTok{, }\StringTok{"CBIO681"}\NormalTok{, }\StringTok{"CBIO703"}\NormalTok{,}
             \StringTok{"CBIO715"}\NormalTok{, }\StringTok{"CBIO725"}\NormalTok{, }\StringTok{"CBIO754"}\NormalTok{,}
             \StringTok{"GIGA"}\NormalTok{)}

\ControlFlowTok{for}\NormalTok{ (batch }\ControlFlowTok{in}\NormalTok{ batches) \{}
\NormalTok{  scp }\OtherTok{\textless{}{-}} \FunctionTok{joinAssays}\NormalTok{(scp,}
                    \AttributeTok{i =} \FunctionTok{grep}\NormalTok{(}\FunctionTok{paste0}\NormalTok{(}\StringTok{"peptide\_"}\NormalTok{, batch), }\FunctionTok{names}\NormalTok{(scp)),}
                    \AttributeTok{name =} \FunctionTok{paste0}\NormalTok{(}\StringTok{"peptides\_"}\NormalTok{, batch))}
\NormalTok{\}}

\NormalTok{scp}
\end{Highlighting}
\end{Shaded}

\begin{verbatim}
## An instance of class QFeatures containing 91 assays:
##  [1] CBIO680_1: SingleCellExperiment with 1862 rows and 3 columns 
##  [2] CBIO680_3: SingleCellExperiment with 1941 rows and 1 columns 
##  [3] CBIO680_4: SingleCellExperiment with 1948 rows and 3 columns 
##  ...
##  [89] peptides_CBIO725: SingleCellExperiment with 4259 rows and 70 columns 
##  [90] peptides_CBIO754: SingleCellExperiment with 6984 rows and 33 columns 
##  [91] peptides_GIGA: SingleCellExperiment with 5138 rows and 26 columns
\end{verbatim}

In this case, 7 new sets are created in the \texttt{scp} object; each of
these new sets combines all the data from each batch.

\hypertarget{peptide-processing}{%
\subsection{Peptide processing}\label{peptide-processing}}

\hypertarget{filtering-of-missing-peptides}{%
\subsubsection{Filtering of missing peptides}\label{filtering-of-missing-peptides}}

Peptides that contain many missing values are not
informative. Peptides with more than 98\% missing data are removed
using the \texttt{filterNA()} function from the \texttt{QFeatures} package. See
\emph{Note 3} for additional recommendations about missing data filtering.

\begin{Shaded}
\begin{Highlighting}[]
\FunctionTok{nrows}\NormalTok{(scp)[}\FunctionTok{grep}\NormalTok{(}\StringTok{"peptides"}\NormalTok{, }\FunctionTok{names}\NormalTok{(scp))]}
\end{Highlighting}
\end{Shaded}

\begin{verbatim}
## peptides_CBIO680 peptides_CBIO681 peptides_CBIO703 peptides_CBIO715 
##             2109             4005             4980             5217 
## peptides_CBIO725 peptides_CBIO754    peptides_GIGA 
##             4259             6984             5138
\end{verbatim}

\begin{Shaded}
\begin{Highlighting}[]
\NormalTok{scp }\OtherTok{\textless{}{-}} \FunctionTok{filterNA}\NormalTok{(scp,}
                \AttributeTok{i =} \FunctionTok{grep}\NormalTok{(}\StringTok{"peptides"}\NormalTok{, }\FunctionTok{names}\NormalTok{(scp)),}
                \AttributeTok{pNA =} \FloatTok{0.98}\NormalTok{)}
\FunctionTok{nrows}\NormalTok{(scp)[}\FunctionTok{grep}\NormalTok{(}\StringTok{"peptides"}\NormalTok{, }\FunctionTok{names}\NormalTok{(scp))]}
\end{Highlighting}
\end{Shaded}

\begin{verbatim}
## peptides_CBIO680 peptides_CBIO681 peptides_CBIO703 peptides_CBIO715 
##             2083             3992             4924             5133 
## peptides_CBIO725 peptides_CBIO754    peptides_GIGA 
##             4190             6822             4991
\end{verbatim}

\hypertarget{normalisation}{%
\subsubsection{Normalisation}\label{normalisation}}

The goal of normalisation is to bring all samples to the same scale
and thus make them comparable\textsuperscript{\protect\hyperlink{ref-cuklina_diagnostics_2021}{25}}. To align
cells' global patterns, we divide all the intensities by the median of
their column. Thus, all the channel intensity distributions are
centred around 1. This is performed using the \texttt{sweep()} function. The
method expects a \texttt{MARGIN}, that defines if the transformation is to be
applied row-wise (1) or column-wise (2), the function (\texttt{FUN}) to apply
and \texttt{STATS}, a vector of values to apply along each column or row (as
defined by \texttt{MARGIN}), in this case, the cell medians. A loop ensures
that normalisation is performed on each ``peptides'' set as computed in
the previous section. Each call to \texttt{sweep()} creates a new set with a
name determined by the \texttt{name} argument. Here, ``\_norm'' is appended at
the end of the original set name.

\begin{Shaded}
\begin{Highlighting}[]
\NormalTok{pep\_assay\_names }\OtherTok{\textless{}{-}} \FunctionTok{names}\NormalTok{(scp)[}\FunctionTok{grep}\NormalTok{(}\StringTok{"peptides\_"}\NormalTok{, }\FunctionTok{names}\NormalTok{(scp))]}

\ControlFlowTok{for}\NormalTok{ (i }\ControlFlowTok{in} \FunctionTok{seq\_along}\NormalTok{(pep\_assay\_names)) \{}
\NormalTok{  scp }\OtherTok{\textless{}{-}} \FunctionTok{sweep}\NormalTok{(scp,}
               \AttributeTok{i =}\NormalTok{ pep\_assay\_names[i],}
               \AttributeTok{MARGIN =} \DecValTok{2}\NormalTok{,}
               \AttributeTok{FUN =} \StringTok{"/"}\NormalTok{,}
               \AttributeTok{STATS =} \FunctionTok{colMedians}\NormalTok{(}\FunctionTok{assay}\NormalTok{(scp[[pep\_assay\_names[i]]]), }\AttributeTok{na.rm =} \ConstantTok{TRUE}\NormalTok{),}
               \AttributeTok{name =} \FunctionTok{paste0}\NormalTok{(pep\_assay\_names[i], }\StringTok{"\_norm"}\NormalTok{))}
\NormalTok{\}}

\NormalTok{scp}
\end{Highlighting}
\end{Shaded}

\begin{verbatim}
## An instance of class QFeatures containing 98 assays:
##  [1] CBIO680_1: SingleCellExperiment with 1862 rows and 3 columns 
##  [2] CBIO680_3: SingleCellExperiment with 1941 rows and 1 columns 
##  [3] CBIO680_4: SingleCellExperiment with 1948 rows and 3 columns 
##  ...
##  [96] peptides_CBIO725_norm: SingleCellExperiment with 4190 rows and 70 columns 
##  [97] peptides_CBIO754_norm: SingleCellExperiment with 6822 rows and 33 columns 
##  [98] peptides_GIGA_norm: SingleCellExperiment with 4991 rows and 26 columns
\end{verbatim}

\hypertarget{log-transformation}{%
\subsubsection{Log transformation}\label{log-transformation}}

Mass spectrometry quantifications have a wide range of values. These
are skewed towards lower values and must be
log-transformed to approximate Gaussian distributions. We perform
log2-transformation on the normalised peptide sets using the
\texttt{logTransform()} method from the \texttt{QFeatures} package.

\begin{Shaded}
\begin{Highlighting}[]
\NormalTok{pep\_assay\_names }\OtherTok{\textless{}{-}} \FunctionTok{names}\NormalTok{(scp)[}\FunctionTok{grep}\NormalTok{(}\StringTok{"peptides\_.*\_norm"}\NormalTok{, }\FunctionTok{names}\NormalTok{(scp))]}

\NormalTok{scp }\OtherTok{\textless{}{-}} \FunctionTok{logTransform}\NormalTok{(scp,}
                    \AttributeTok{base =} \DecValTok{2}\NormalTok{,}
                    \AttributeTok{i =}\NormalTok{ pep\_assay\_names,}
                    \AttributeTok{name =} \FunctionTok{paste0}\NormalTok{(pep\_assay\_names, }\StringTok{"\_log"}\NormalTok{))}
\end{Highlighting}
\end{Shaded}

Similarly to \texttt{sweep()}, \texttt{logTransform()} creates new sets that are
named by appending ``\_log'' to the original names.

\hypertarget{peptide-to-protein-aggregation}{%
\subsubsection{Peptide to protein aggregation}\label{peptide-to-protein-aggregation}}

Similarly to aggregating PSMs into peptides, peptides are aggregated
into proteins using the \texttt{aggregateFeatures()} function. The \texttt{fcol}
argument names the feature variable to use for aggregation, in this
case, the protein accessions defined in \texttt{"proteins"}. The protein
inference method is therefore implicitly left to the pre-processing
algorithm. The \texttt{fun} argument provides the function that will
aggregate the peptide quantitative data. Here we use the median,
complementary informations about protein summarisation can be found in
\emph{Note 4}.

\begin{Shaded}
\begin{Highlighting}[]
\NormalTok{pep\_assay\_names }\OtherTok{\textless{}{-}} \FunctionTok{names}\NormalTok{(scp)[}\FunctionTok{grep}\NormalTok{(}\StringTok{"peptides\_.*\_norm\_log"}\NormalTok{, }\FunctionTok{names}\NormalTok{(scp))]}

\NormalTok{scp }\OtherTok{\textless{}{-}} \FunctionTok{aggregateFeatures}\NormalTok{(scp,}
                         \AttributeTok{i =}\NormalTok{ pep\_assay\_names,}
                         \AttributeTok{fcol =} \StringTok{"proteins"}\NormalTok{,}
                         \AttributeTok{fun =}\NormalTok{ colMedians, }\AttributeTok{na.rm =} \ConstantTok{TRUE}\NormalTok{,}
                         \AttributeTok{name =} \FunctionTok{sub}\NormalTok{(}\StringTok{"peptides"}\NormalTok{, }\StringTok{"proteins"}\NormalTok{, pep\_assay\_names))}
\end{Highlighting}
\end{Shaded}

\begin{figure}
\centering
\includegraphics{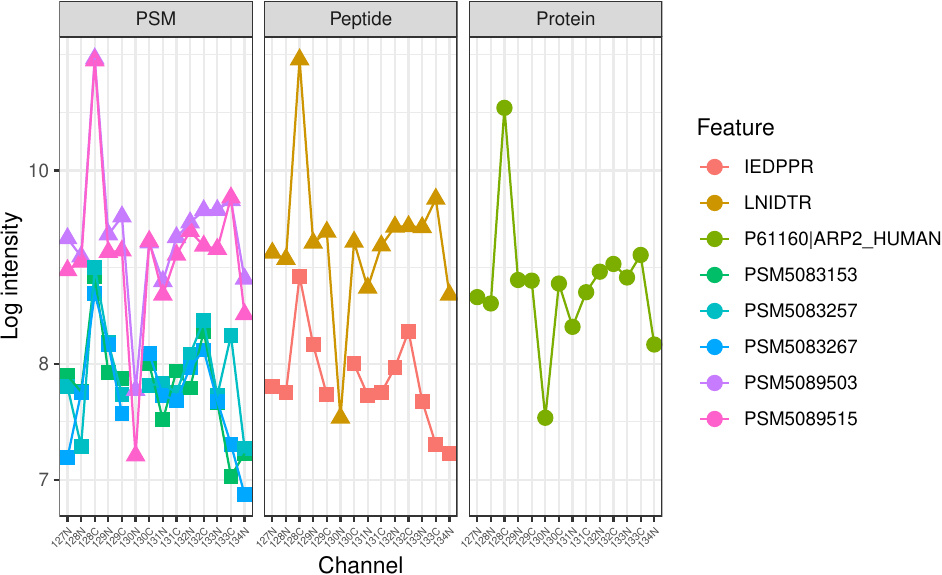}
\caption{\label{fig:fullAggregationExample}Example of aggregation of PSM data into peptide and protein data. PSM intensities for each channel are plotted on the left panel. PSMs matched to the peptide `IEDPPR' are represented by the line and squares while PSMs matched to the peptide `LNIDTR' are represented by the line and triangles. Intensities of the peptides resulting from the aggregation of the PSMs are plotted on the middle panel. Intensities of the protein resulting from the aggregation of the peptides are plotted on the right panel.}
\end{figure}

\textbf{Figure \ref{fig:fullAggregationExample}} illustrates the effect of
aggregation from PSMs to peptides and proteins.

\hypertarget{protein-processing}{%
\subsection{Protein processing}\label{protein-processing}}

\hypertarget{imputation}{%
\subsubsection{Imputation}\label{imputation}}

Imputation consists of replacing missing values with predicted values. These
imputed values are computed from the observed values. One of the most
commonly used algorithms is the k-Nearest Neighbours (KNN)
algorithm. KNN infers values from features showing similar expression
patterns, called neighbours, across samples. Working with a complete
dataset unlocks many computational tools that break upon the presence
of missing values. However, imputation of missing values can lead to
biased estimates\textsuperscript{\protect\hyperlink{ref-obrien_effects_2018}{26},\protect\hyperlink{ref-goeminne_msqrob_2020}{27}},
especially for data with high proportions of missing values such as
SCP\textsuperscript{\protect\hyperlink{ref-vanderaa_revisiting_2023}{28}}. Therefore, we recommend avoiding
imputation when possible.

If needed, imputation can easily be done with the \texttt{impute()}
function. \texttt{impute()} needs the index of the sets to impute, here the
normalised and log-transformed protein sets. The method used is the
KNN algorithm with 3 nearest neighbours (\texttt{k}). The \texttt{rowmax} and
\texttt{colmax} arguments limit the maximum percentage of missing data allowed in
any row and column, respectively. We set both to 1 to allow any
proportion of missing value. We name the imputed sets by substituting
the ``norm\_log'' suffixes with ``imptd''.

\begin{Shaded}
\begin{Highlighting}[]
\FunctionTok{table}\NormalTok{(}\FunctionTok{is.na}\NormalTok{(}\FunctionTok{assay}\NormalTok{(scp[[}\StringTok{"proteins\_CBIO680\_norm\_log"}\NormalTok{]])))}
\end{Highlighting}
\end{Shaded}

\begin{verbatim}
## 
## FALSE  TRUE 
##  4718  1561
\end{verbatim}

\begin{Shaded}
\begin{Highlighting}[]
\NormalTok{prot\_assay\_names }\OtherTok{\textless{}{-}} \FunctionTok{names}\NormalTok{(scp)[}\FunctionTok{grep}\NormalTok{(}\StringTok{"proteins.*\_norm\_log"}\NormalTok{, }\FunctionTok{names}\NormalTok{(scp))]}

\NormalTok{scp }\OtherTok{\textless{}{-}} \FunctionTok{impute}\NormalTok{(scp,}
              \AttributeTok{i =}\NormalTok{ prot\_assay\_names,}
              \AttributeTok{method =} \StringTok{"knn"}\NormalTok{,}
              \AttributeTok{k =} \DecValTok{3}\NormalTok{, }\AttributeTok{rowmax =} \DecValTok{1}\NormalTok{, }\AttributeTok{colmax=} \DecValTok{1}\NormalTok{,}
              \AttributeTok{name =} \FunctionTok{sub}\NormalTok{(}\StringTok{"norm\_log"}\NormalTok{, }\StringTok{"imptd"}\NormalTok{, prot\_assay\_names))}
\end{Highlighting}
\end{Shaded}

Imputed sets do not contain any missing values anymore.

\begin{Shaded}
\begin{Highlighting}[]
\FunctionTok{any}\NormalTok{(}\FunctionTok{is.na}\NormalTok{(}\FunctionTok{assay}\NormalTok{(scp[[}\StringTok{"proteins\_CBIO680\_imptd"}\NormalTok{]])))}
\end{Highlighting}
\end{Shaded}

\begin{verbatim}
## [1] FALSE
\end{verbatim}

\hypertarget{batch-correction}{%
\subsubsection{Batch correction}\label{batch-correction}}

Data need to be corrected for batch effects. Batch effects are caused
by technical fluctuations occurring during different MS runs. Since only
one to a small number of multiplexed single cells can be acquired at
once, batch effects are unavoidable.

\begin{figure}
\centering
\includegraphics{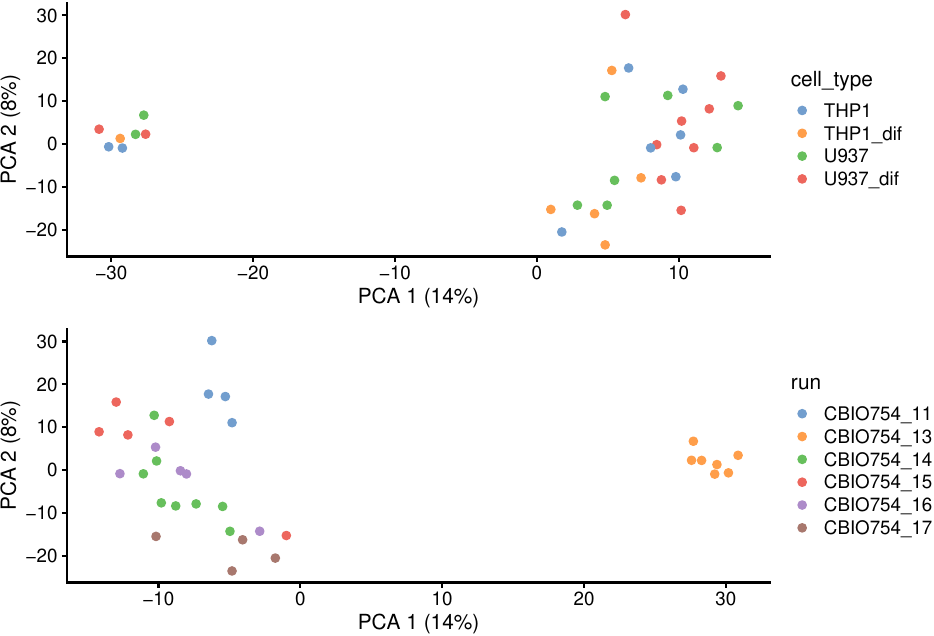}
\caption{\label{fig:pcaBatchEffects}Principal component analysis of CBIO754 cells without batch correction. Cells are coloured based on their cell type on top and their MS run on the bottom. Cells cluster based on their acquisition batch rather than their cell type.}
\end{figure}

When performing a principal component analysis (PCA) at this stage, we
can see that cells cluster together based on the MS run they were
acquired in (technical variability) rather than based on their cell
type (biological variability) (\textbf{Figure
\ref{fig:pcaBatchEffects}}). Additional technical variability is
induced by the channel used for each cell. Batch correction allows
removing this technical variability without altering biological
variability. Note that PCA will be described later, in section
\ref{dimred}.

A loop is performed to repeat the batch correction on each set, both
imputed and not-imputed peptide-level data.

\begin{itemize}
\tightlist
\item
  We extract the set on which the batch correction is performed using
  the \texttt{getWithColData()} function that returns an annotated
  \texttt{SingleCellExperiment} object.
\item
  The \texttt{removeBatchEffect()} function from the \texttt{limma} package is used
  to perform batch correction on the \texttt{assay}. \texttt{removeBatchEffect()}
  uses two types of arguments: \texttt{group} to define the variable to be
  preserved, here the \texttt{cell\_type}, and \texttt{batch} and \texttt{batch2} to define
  the technical variables to be corrected, here \texttt{run} and \texttt{channel}.
\item
  The batch corrected set is added to the \texttt{QFeatures} object and a
  link between the corrected set and the original one is created to
  traceback parent and child assays.
\end{itemize}

Note that batches \texttt{CBIO680} and \texttt{CBIO681} cannot be batch corrected
properly as the \texttt{channel} variable is confounded with the \texttt{cell\_type}
(see \emph{Note 5}). It is important to randomise cells across acquisition
runs and channels\textsuperscript{\protect\hyperlink{ref-gatto_initial_2023}{23}}.

\begin{Shaded}
\begin{Highlighting}[]
\ControlFlowTok{for}\NormalTok{ (i }\ControlFlowTok{in} \FunctionTok{grep}\NormalTok{(}\StringTok{"norm\_log|imptd"}\NormalTok{, }\FunctionTok{names}\NormalTok{(scp))) \{}
  \DocumentationTok{\#\# Extract set}
\NormalTok{  sce }\OtherTok{\textless{}{-}} \FunctionTok{getWithColData}\NormalTok{(scp, }\FunctionTok{names}\NormalTok{(scp)[i])}
  \DocumentationTok{\#\# Batch correct assay}
  \FunctionTok{assay}\NormalTok{(sce) }\OtherTok{\textless{}{-}}
    \FunctionTok{removeBatchEffect}\NormalTok{(}\FunctionTok{assay}\NormalTok{(sce), }\AttributeTok{group =}\NormalTok{ sce}\SpecialCharTok{$}\NormalTok{cell\_type,}
                      \AttributeTok{batch =}\NormalTok{ sce}\SpecialCharTok{$}\NormalTok{run, }\AttributeTok{batch2 =}\NormalTok{ sce}\SpecialCharTok{$}\NormalTok{channel)}
  \DocumentationTok{\#\# Name and add batch{-}corrected assay}
\NormalTok{  scp }\OtherTok{\textless{}{-}} \FunctionTok{addAssay}\NormalTok{(scp,}
                  \AttributeTok{y =}\NormalTok{ sce,}
                  \AttributeTok{name =} \FunctionTok{sub}\NormalTok{(}\StringTok{"\_norm\_log|mptd"}\NormalTok{, }\StringTok{"\_batchC"}\NormalTok{, }\FunctionTok{names}\NormalTok{(scp)[i]))}
  \DocumentationTok{\#\# Add link between batch corrected and original assay}
\NormalTok{  scp }\OtherTok{\textless{}{-}} \FunctionTok{addAssayLinkOneToOne}\NormalTok{(scp,}
                              \AttributeTok{from =} \FunctionTok{names}\NormalTok{(scp)[i],}
                              \AttributeTok{to =} \FunctionTok{sub}\NormalTok{(}\StringTok{"\_norm\_log|mptd"}\NormalTok{, }\StringTok{"\_batchC"}\NormalTok{, }\FunctionTok{names}\NormalTok{(scp)[i]))}
\NormalTok{\}}
\end{Highlighting}
\end{Shaded}

\begin{Shaded}
\begin{Highlighting}[]
\NormalTok{scp}
\end{Highlighting}
\end{Shaded}

\begin{verbatim}
## An instance of class QFeatures containing 140 assays:
##  [1] CBIO680_1: SingleCellExperiment with 1862 rows and 3 columns 
##  [2] CBIO680_3: SingleCellExperiment with 1941 rows and 1 columns 
##  [3] CBIO680_4: SingleCellExperiment with 1948 rows and 3 columns 
##  ...
##  [138] proteins_CBIO725_i_batchC: SingleCellExperiment with 1360 rows and 70 columns 
##  [139] proteins_CBIO754_i_batchC: SingleCellExperiment with 1801 rows and 33 columns 
##  [140] proteins_GIGA_i_batchC: SingleCellExperiment with 1629 rows and 26 columns
\end{verbatim}

\hypertarget{dimred}{%
\subsection{Dimensionality reduction}\label{dimred}}

We will demonstrate 2 approaches to reduce dimensions using principal
component analysis (PCA), one where missing values are retained
(Nonlinear Iterative Partial Least Squares, NIPALS), and one where
missing values are imputed (Singular value decomposition, SVD).

\hypertarget{nonlinear-iterative-partial-least-squares}{%
\subsubsection{Nonlinear Iterative Partial Least Squares}\label{nonlinear-iterative-partial-least-squares}}

Principal component analyses are run on each batch corrected set using
the \texttt{pca()} function from \texttt{pcaMethods}. We chose the ``NIPALS'' method
as this algorithm can handle missing values. We build a loop to
perform dimensionality reduction on all batch corrected sets (both
with and without missing values). Note that it's not necessary to
perform the NIPALS method on imputed sets since we use the NIPALS
method precisely to avoid imputation, but we do so nonetheless for
illustrative purposes. The quantitative matrix (\texttt{assay}) of the set is
extracted and its missing values are encoded in the supported format
(from \texttt{NaN} to \texttt{NA}). The quantitative \texttt{assay} is transposed before
the PCA is performed, so that rows represent cells and columns
represent features as expected by \texttt{pcaMethods}. Dimensionality
reduction results are stored in the corresponding
\texttt{SingleCellExperiment} set within the \texttt{scp} object in the \texttt{reducedDim}
slot (see \textbf{Figure \ref{fig:seqf}}).

\begin{Shaded}
\begin{Highlighting}[]
\ControlFlowTok{for}\NormalTok{ (i }\ControlFlowTok{in} \FunctionTok{grep}\NormalTok{(}\StringTok{"batchC"}\NormalTok{, }\FunctionTok{names}\NormalTok{(scp))) \{}
\NormalTok{  nipals\_res }\OtherTok{\textless{}{-}}
      \DocumentationTok{\#\# Extract assay}
      \FunctionTok{assay}\NormalTok{(scp[[i]]) }\SpecialCharTok{|\textgreater{}}
      \FunctionTok{as.data.frame}\NormalTok{() }\SpecialCharTok{|\textgreater{}}
      \DocumentationTok{\#\# Encode missing values}
      \FunctionTok{mutate\_all}\NormalTok{(}\SpecialCharTok{\textasciitilde{}}\FunctionTok{ifelse}\NormalTok{(}\FunctionTok{is.nan}\NormalTok{(.), }\ConstantTok{NA}\NormalTok{, .)) }\SpecialCharTok{|\textgreater{}}
      \DocumentationTok{\#\# Transpose}
      \FunctionTok{t}\NormalTok{() }\SpecialCharTok{|\textgreater{}}
      \DocumentationTok{\#\# PCA}
\NormalTok{      pcaMethods}\SpecialCharTok{::}\FunctionTok{pca}\NormalTok{(}\AttributeTok{method=}\StringTok{"nipals"}\NormalTok{, }\AttributeTok{nPcs =} \DecValTok{2}\NormalTok{)}

  \FunctionTok{reducedDim}\NormalTok{(scp[[i]], }\StringTok{"NIPALS"}\NormalTok{) }\OtherTok{\textless{}{-}}\NormalTok{ pcaMethods}\SpecialCharTok{::}\FunctionTok{scores}\NormalTok{(nipals\_res)}
\NormalTok{\}}
\end{Highlighting}
\end{Shaded}

Reduced dimensions can then be accessed using the \texttt{reducedDim()}
function and plotted with the \texttt{plotReducedDim()} function.

\begin{Shaded}
\begin{Highlighting}[]
\FunctionTok{head}\NormalTok{(}\FunctionTok{reducedDim}\NormalTok{(scp[[}\StringTok{"proteins\_CBIO703\_batchC"}\NormalTok{]], }\StringTok{"NIPALS"}\NormalTok{))}
\end{Highlighting}
\end{Shaded}

\begin{verbatim}
##                        PC1       PC2
## CBIO703_2_130C -2.76973573 -1.243693
## CBIO703_2_132C -0.38621841  3.655280
## CBIO703_2_133N  0.08611501  3.445896
## CBIO703_2_133C  0.12780095  4.735788
## CBIO703_3_128C -8.98370688 -2.023013
## CBIO703_3_131N  5.56700122 -3.223104
\end{verbatim}

\begin{Shaded}
\begin{Highlighting}[]
\NormalTok{NIPALS\_CBIO703  }\OtherTok{\textless{}{-}}
  \FunctionTok{plotReducedDim}\NormalTok{(scp[[}\StringTok{"proteins\_CBIO703\_batchC"}\NormalTok{]],}
                 \AttributeTok{dimred =} \StringTok{"NIPALS"}\NormalTok{,}
                 \AttributeTok{color\_by =} \StringTok{"cell\_type"}\NormalTok{,}
                 \AttributeTok{point\_alpha =} \DecValTok{1}\NormalTok{)}

\NormalTok{NIPALS\_CBIO754  }\OtherTok{\textless{}{-}}
  \FunctionTok{plotReducedDim}\NormalTok{(scp[[}\StringTok{"proteins\_CBIO754\_batchC"}\NormalTok{]],}
                 \AttributeTok{dimred =} \StringTok{"NIPALS"}\NormalTok{,}
                 \AttributeTok{color\_by =} \StringTok{"cell\_type"}\NormalTok{,}
                 \AttributeTok{point\_alpha =} \DecValTok{1}\NormalTok{)}
\end{Highlighting}
\end{Shaded}

\begin{figure}
\centering
\includegraphics{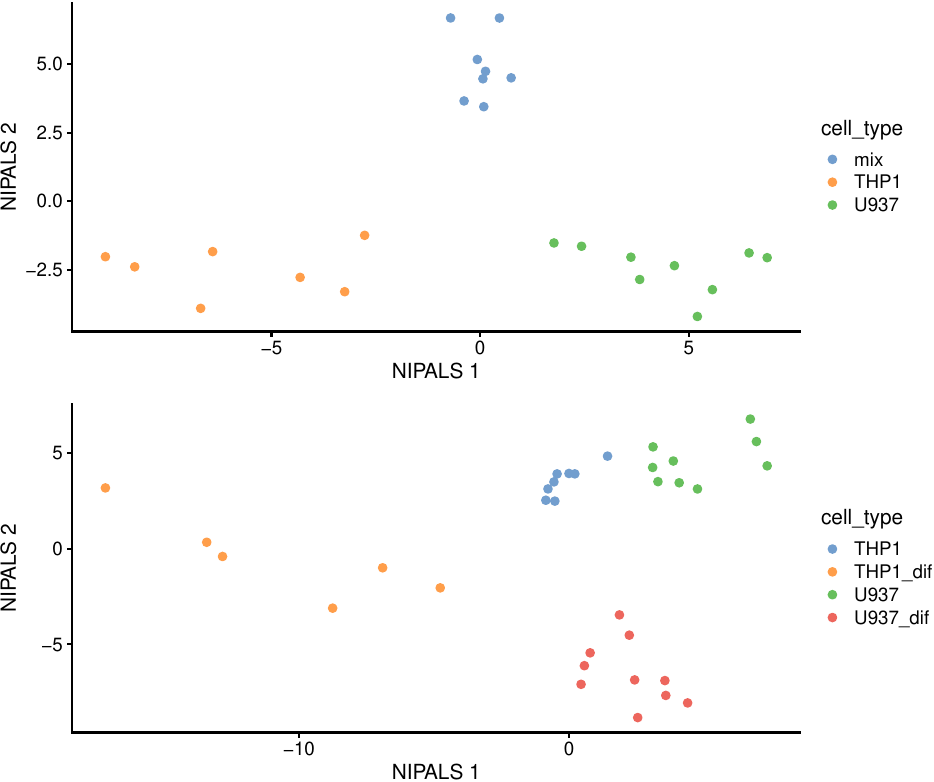}
\caption{\label{fig:plotCombining}NIPALS PCA of CBIO703 and CBIO754 cells. Cells from two batches (CBIO753 on top and CBIO754 on the bottom) are coloured based on their cell types.}
\end{figure}

Below, we combine 2 PCA plots using the \texttt{patchwork} package (\textbf{Figure
\ref{fig:plotCombining}}).

\begin{Shaded}
\begin{Highlighting}[]
\NormalTok{NIPALS\_CBIO703 }\SpecialCharTok{/}\NormalTok{ NIPALS\_CBIO754}
\end{Highlighting}
\end{Shaded}

Using this workflow, single cells cluster well together on a PCA based
on their cell type (\textbf{Figure \ref{fig:plotCombining}}). This is the
case for our 2 designs: THP1/U937/mix and THP1/THP1\_dif/U937/U937\_dif.

\hypertarget{singular-value-decomposition}{%
\subsubsection{Singular value decomposition}\label{singular-value-decomposition}}

The singular value decomposition (SVD) method is more commonly used
for PCA. However, it cannot handle missing values and thus requires
imputed sets. The \texttt{runPCA()} function from the \texttt{scater} package is an
easy way to perform SVD PCA on the imputed \texttt{SummarizedExperiment} sets
within the \texttt{QFeatures} object. Reduced dimensions are directly stored
in the \texttt{ReducedDim} slot with the name provided by the \texttt{name}
argument.

\begin{Shaded}
\begin{Highlighting}[]
\ControlFlowTok{for}\NormalTok{ (i }\ControlFlowTok{in} \FunctionTok{grep}\NormalTok{(}\StringTok{"\_i\_batchC"}\NormalTok{, }\FunctionTok{names}\NormalTok{(scp))) \{}
\NormalTok{  scp[[i]] }\OtherTok{\textless{}{-}} \FunctionTok{runPCA}\NormalTok{(scp[[i]],}
                     \AttributeTok{ncomponents =} \DecValTok{5}\NormalTok{,}
                     \AttributeTok{ntop =} \ConstantTok{Inf}\NormalTok{,}
                     \AttributeTok{scale =} \ConstantTok{TRUE}\NormalTok{,}
                     \AttributeTok{exprs\_values =} \DecValTok{1}\NormalTok{,}
                     \AttributeTok{name =} \StringTok{"SVD"}\NormalTok{)}
\NormalTok{\}}
\end{Highlighting}
\end{Shaded}

Since we computed both SVD and NIPALS on the imputed
\texttt{SingleCellExperiment} sets, they now have 2 elements in the
\texttt{ReducedDim} slot, one for NIPALS and one for SVD.

\begin{Shaded}
\begin{Highlighting}[]
\NormalTok{scp[[}\StringTok{"proteins\_CBIO754\_i\_batchC"}\NormalTok{]]}
\end{Highlighting}
\end{Shaded}

\begin{verbatim}
## class: SingleCellExperiment 
## dim: 1801 33 
## metadata(0):
## assays(2): assay aggcounts
## rownames(1801):
##   sp|A0A096LP55|QCR6L_HUMAN;sp|A0A096LP55|QCR6L_HUMAN;sp|P07919|QCR6_HUMAN;sp|P07919|QCR6_HUMAN
##   sp|A0A0B4J2A2|PAL4C_HUMAN;sp|A0A075B767|PAL4H_HUMAN;sp|P0DN37|PAL4G_HUMAN;sp|F5H284|PAL4D_HUMAN;sp|Q9Y536|PAL4A_HUMAN;sp|P62937|PPIA_HUMAN;sp|P0DN26|PAL4F_HUMAN;sp|A0A075B759|PAL4E_HUMAN
##   ... sp|Q9Y6N5|SQOR_HUMAN
##   tr|A0A8I5KQE6|A0A8I5KQE6_HUMAN;sp|P08865|RSSA_HUMAN
## rowData names(11): proteins num_proteins ... chimeric .n
## colnames(33): CBIO754_11_128N CBIO754_11_131C ... CBIO754_17_133C
##   CBIO754_17_134N
## colData names(9): run channel ... medianCV count
## reducedDimNames(2): NIPALS SVD
## mainExpName: NULL
## altExpNames(0):
\end{verbatim}

SVD reduced dimensions can be accessed using the \texttt{reducedDim()}
function or plotted using \texttt{plotReducedDim()}, by specifying the ``SVD''
name.

\begin{Shaded}
\begin{Highlighting}[]
\FunctionTok{head}\NormalTok{(}\FunctionTok{reducedDim}\NormalTok{(scp[[}\StringTok{"proteins\_CBIO754\_i\_batchC"}\NormalTok{, }\StringTok{"SVD"}\NormalTok{]]))}
\end{Highlighting}
\end{Shaded}

\begin{verbatim}
##                       PC1       PC2
## CBIO754_11_128N -6.190958 -7.184588
## CBIO754_11_131C -2.367707  7.621617
## CBIO754_11_132C  7.620063  3.247906
## CBIO754_11_133C -1.943787 -1.093237
## CBIO754_13_127N -2.808502  6.770368
## CBIO754_13_128C  8.303595  1.147066
\end{verbatim}

\begin{Shaded}
\begin{Highlighting}[]
\NormalTok{svd\_CBIO703 }\OtherTok{\textless{}{-}}
  \FunctionTok{plotReducedDim}\NormalTok{(scp[[}\StringTok{"proteins\_CBIO703\_i\_batchC"}\NormalTok{]],}
                 \AttributeTok{dimred =} \StringTok{"SVD"}\NormalTok{,}
                 \AttributeTok{color\_by =} \StringTok{"cell\_type"}\NormalTok{,}
                 \AttributeTok{point\_alpha =} \DecValTok{1}\NormalTok{)}

\NormalTok{svd\_CBIO754 }\OtherTok{\textless{}{-}}
  \FunctionTok{plotReducedDim}\NormalTok{(scp[[}\StringTok{"proteins\_CBIO754\_i\_batchC"}\NormalTok{]],}
                 \AttributeTok{dimred =} \StringTok{"SVD"}\NormalTok{,}
                 \AttributeTok{color\_by =} \StringTok{"cell\_type"}\NormalTok{,}
                 \AttributeTok{point\_alpha =} \DecValTok{1}\NormalTok{)}

\NormalTok{svd\_CBIO703 }\SpecialCharTok{/}\NormalTok{ svd\_CBIO754}
\end{Highlighting}
\end{Shaded}

\begin{figure}
\centering
\includegraphics{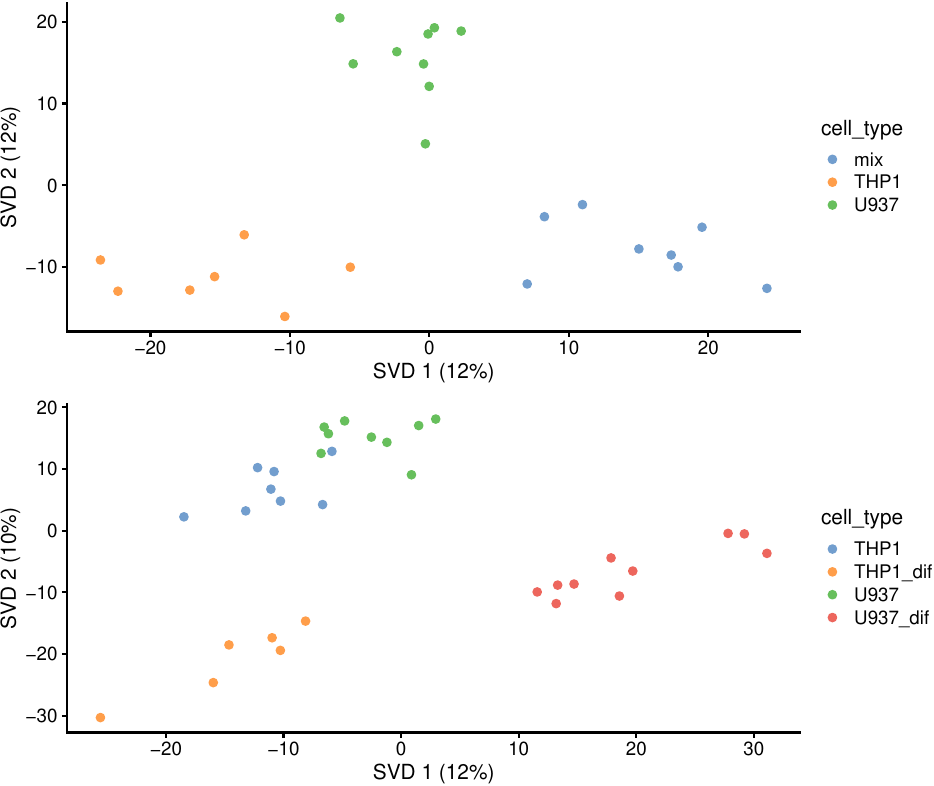}
\caption{\label{fig:plotPca}SVD PCA of CBIO703 and CBIO754 cells. Cells from two batches (CBIO753 on top and CBIO754 on the bottom) are coloured based on their cell types.}
\end{figure}

Despite the imputation, we observe similar results to NIPALS (\textbf{Figure
\ref{fig:plotPca}}).

\hypertarget{downstream-analysis}{%
\subsection{Downstream analysis}\label{downstream-analysis}}

The fully processed sets, i.e.~batch-corrected protein sets, are ready
for further analysis. We have already mentioned one type of downstream
analysis in the previous section with dimensionality reduction, but
many other analyses can be carried out to reveal biological insights
from single-cell data. While we will not go into details on how to
perform downstream analysis as this is not the scope of this protocol,
below we suggest tools for some of the approaches commonly applied to
SCP data.

UMAP was used by Schoof et al.~(2021)\textsuperscript{\protect\hyperlink{ref-schoof_quantitative_2021}{29}} and Petrosius et al.~(2023)\textsuperscript{\protect\hyperlink{ref-petrosius_exploration_2023}{30}} for dimensionality reduction. UMAP and
t-SNE are non-linear dimensionality reduction techniques focusing on a
specific neighbourhood rather than distances between cells. They can
be used as an alternative to PCA. We can easily perform UMAP and t-SNE
using \texttt{runUMAP()} and \texttt{runTSNE()} functions from the \texttt{scater} package\textsuperscript{\protect\hyperlink{ref-McCarthy_scater_2017}{15}}, similarly to \texttt{runPCA()}.

Clustering is an unsupervised learning procedure that is used to
empirically define groups of cells with similar expression
profiles. Schoof et al.~(2021)\textsuperscript{\protect\hyperlink{ref-schoof_quantitative_2021}{29}} built a KNN graph and used Leiden
community detection to perform clustering. This can be carried out
with the \texttt{clusterCells()} function from the \texttt{scran} package\textsuperscript{\protect\hyperlink{ref-Lun_scran_2016}{31}}.

For differential expression analysis (DEA), many approaches use the
t-test\textsuperscript{\protect\hyperlink{ref-schoof_quantitative_2021}{29},\protect\hyperlink{ref-liang_fully_2021}{32},\protect\hyperlink{ref-brunner_ultrahigh_2022}{33}}. The \texttt{t.test()} function from the base
package is the simplest option. Alternatively, linear models as
provided by the \texttt{limma} package\textsuperscript{\protect\hyperlink{ref-Ritchie_limma_2015}{14}} offer more
flexible approaches.

Protein set enrichment analysis (PSEA) was performed by Leduc et al.~(2022)\textsuperscript{\protect\hyperlink{ref-leduc_exploring_2022}{1}} to identify sets of proteins of interest,
i.e.~proteins with similar functions or involved in the same process,
that are enriched for differential expression. Functions like
\texttt{enrichGO()} from the \texttt{clusterProfiler} package\textsuperscript{\protect\hyperlink{ref-Wu_clusterProfiler_2021}{34},\protect\hyperlink{ref-Yu_clusterProfiler_2012}{35}} can be used for
enrichment analyses and visualisation thereof.

Trajectory inference is used to arrange cells based on their
progression through a dynamic process like cell differentiation or
cell cycle. Schoof et al.~(2021)\textsuperscript{\protect\hyperlink{ref-schoof_quantitative_2021}{29}} used diffusion pseudo-time
(DPT). DPT can be plotted using the \texttt{DiffusionMap()} function from the
destiny package\textsuperscript{\protect\hyperlink{ref-Angerer_destiny_2015}{36}}. Zhu et al.~(2019)\textsuperscript{\protect\hyperlink{ref-zhu_single-cell_2019}{37}} used the
\texttt{CellTrails} package\textsuperscript{\protect\hyperlink{ref-Ellwanger_CellTrails_2018}{38}}.

The \texttt{SingleCellExperiment} class, used as part of the \texttt{scp} pipeline,
provides direct compatibility with all of these tools.

\hypertarget{notes}{%
\section{Notes}\label{notes}}

\begin{figure}
\centering
\includegraphics{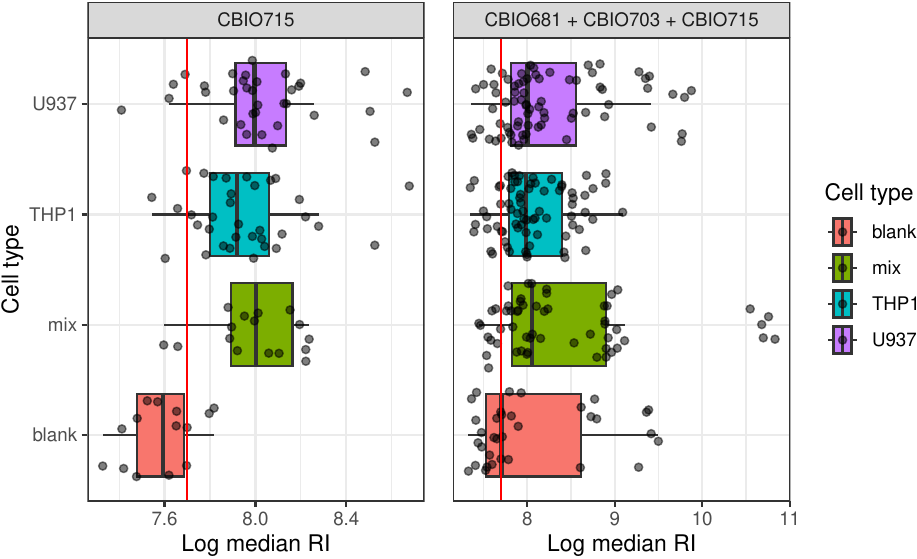}
\caption{\label{fig:figNote1}log median RI for one batch vs log median RI for grouped batches.}
\end{figure}

\begin{enumerate}
\def\labelenumi{\arabic{enumi}.}
\tightlist
\item
  Metrics used for quality control can be heavily influenced by
  non-biological parameters like instrument types, instrument
  performances and experimental design. Based on our experience, we
  recommend performing quality control on every acquisition batch
  rather than grouping every run. \textbf{Figure \ref{fig:figNote1}}
  shows on the left median RI distribution for one batch, and on the
  right median RI distribution for the 3 batches. The three batches
  were run on the same mass spectrometer and they contain the same
  cell types. While the separation between blanks and cells is clear
  in CBIO715 alone, this separation becomes blurred when we combine
  the 3 batches.
\end{enumerate}

\begin{figure}
\centering
\includegraphics{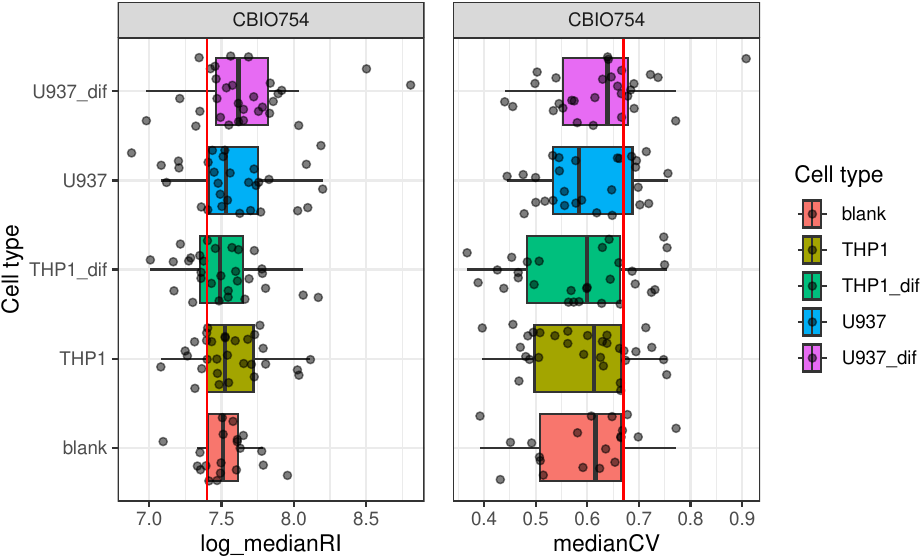}
\caption{\label{fig:noteQC}median CV distribution per cell type for batch CBIO754.}
\end{figure}

\begin{enumerate}
\def\labelenumi{\arabic{enumi}.}
\setcounter{enumi}{1}
\tightlist
\item
  Blanks do not always exhibit different distributions than single-cell
  samples. In our dataset, it is for example the case for batch
  \texttt{CBIO754} (\textbf{Figure \ref{fig:noteQC}}). In this situation, it
  might be necessary to set an arbitrary threshold to remove single
  cells with the lowest median RI and highest median CV without
  considering blanks. However, great care needs to be
  taken in downstream analyses to ensure that the data remain useful.
\end{enumerate}

\begin{figure}
\centering
\includegraphics{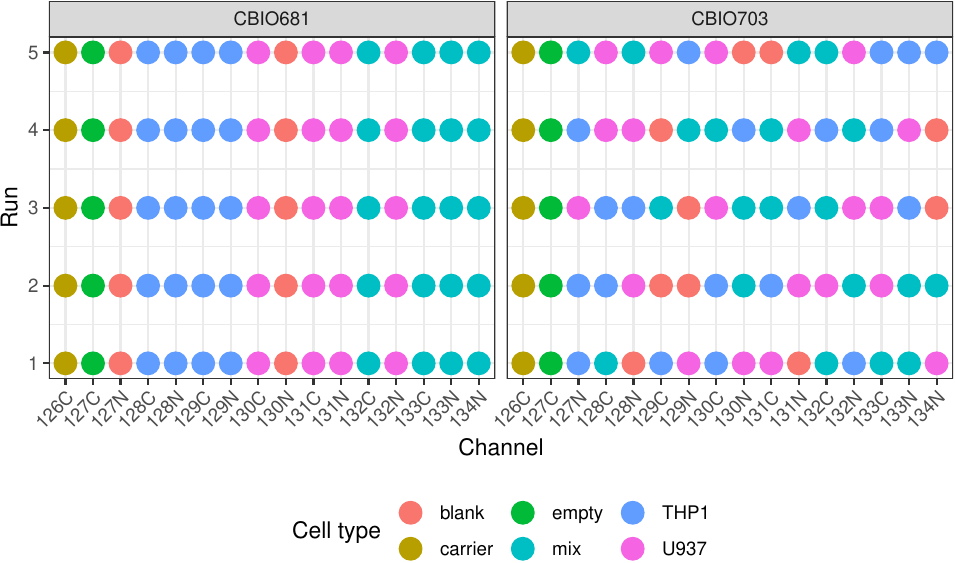}
\caption{\label{fig:figDesign}Non-randomised design vs randomised design. Each dot represent a cell. On the left panel, cells in the same channel have the same cell type across all runs. On the right panel, cell with the same cell type are randomised across the channels.}
\end{figure}

\begin{enumerate}
\def\labelenumi{\arabic{enumi}.}
\setcounter{enumi}{2}
\item
  We advise awareness about the proportion of peptides getting
  removed at the NA filtering step as peptide missingness can widely
  vary depending on the number of samples in the dataset\textsuperscript{\protect\hyperlink{ref-vanderaa_revisiting_2023}{28}}. Using a low \texttt{pNA} can easily remove
  most of your peptides if a large dataset is used.
\item
  In bottum-up proteomics, summarisation of peptide intensities
  toward protein abundances are impacted by various factors. First,
  different peptides from the same protein often have very distinct
  physio-chemical properties, leading to large differences in their
  MS1 intensities even though these peptides are of similar abundance\textsuperscript{\protect\hyperlink{ref-sticker_robust_2020}{39}}. Second, with data dependent acquisition
  (DDA), only those peptides with the highest MS1 intensities within
  a certain retention window are selected for fragmentation. As a
  result, peptide selection varies not only with abundance, but also
  with context, and therefore varies stochastically between runs\textsuperscript{\protect\hyperlink{ref-tu_systematic_2014}{40}}. These sources of missingness imply that the
  peptide data can either be missing at random (MAR) or missing not
  at random (MNAR)\textsuperscript{\protect\hyperlink{ref-lazar_accounting_2016}{41}}. Under these conditions,
  peptide to protein aggregation by summing peptide intensities
  should be avoided, since missing peptides will be considered as
  missing because they are not present in the sample (MNAR), which is
  not necessarily the case. Indeed, summing peptide intensities would
  result in an implicit imputation of missing data by 0, which should
  be avoided due to the complexity of missing data specific to
  bottom-up proteomics data\textsuperscript{\protect\hyperlink{ref-Kong2022-wp}{24}}. In this protocol we used the median to
  summarise the peptide data. Functions using robust statistical
  methods like \texttt{robustSummary()} from the \texttt{MsCoreUtils} package\textsuperscript{\protect\hyperlink{ref-goeminne_msqrob_2020}{27},\protect\hyperlink{ref-rainer_modular_2022}{42}} take more time to
  compute but are well suited for this kind of summarisation.
\item
  Experimental design should be thoroughly planned so that biological
  and technical variability can be disambiguated. Acquisition batches
  \texttt{CBIO680} and \texttt{CBIO681} follow a design where channels contain the
  same cell type across all runs (\textbf{Figure \ref{fig:figDesign}},
  left). In doing so, biological factors (cell type) and technical
  factors (channel) are confounded and batch effects cannot be
  modelled and corrected. Proper cell type randomisation across
  channels is shown in \textbf{Figure \ref{fig:figDesign}}, on the right.
\end{enumerate}

\hypertarget{computational-requirements}{%
\section{Computational requirements}\label{computational-requirements}}

The median time to run the complete workflow was 5.34
minutes. Detailed timings for steps taking over 5 seconds are shown in
\textbf{Table \ref{tab:timetab}}. The processing used a total 3.65 GB
of memory. Benchmarks were repeated 5 times on a virtual machine
running 1 CPU Epyc MILAN 7313 16 Cores @ 3.0Ghz (limited to 8 cores on
the VM) and 64 Gb of RAM DDR4 ECC.

\begin{table}

\caption{\label{tab:timetab}Computation time used for each step of the workflow taking more than 5 seconds.}
\centering
\begin{tabular}[t]{llr}
\toprule
  & Workflow step & Median time (s)\\
\midrule
1 & PSM to peptide aggregation & 57.25\\
2 & Building of the QFeatures object & 45.05\\
3 & Highlighting of chimeric spectra & 26.35\\
4 & Removal of carrier and empty channels & 25.93\\
5 & NIPALS computation & 22.15\\
\addlinespace
6 & Batch correction & 19.07\\
7 & Reading of rds object & 18.19\\
8 & Subsetting of filtered samples & 11.46\\
9 & Substitution of 0s by NAs & 10.19\\
10 & Peptide to protein aggregation & 9.96\\
\addlinespace
11 & Joining of peptide batches & 9.05\\
12 & Imputation & 8.64\\
13 & Peptide log-transformation & 7.29\\
14 & Removal of blanks & 7.06\\
15 & SVD computation & 7.00\\
\addlinespace
16 & Peptide normalisation & 6.66\\
17 & SCR computation & 6.36\\
\bottomrule
\end{tabular}
\end{table}

\hypertarget{session-information}{%
\section{Session information}\label{session-information}}

A complete list and version of R and packages used to run this
complete analysis and produce the results is provided below.

\begin{itemize}\raggedright
  \item R version 4.3.1 Patched (2023-07-10 r84676), \verb|x86_64-pc-linux-gnu|
  \item Locale: \verb|LC_CTYPE=en_US.UTF-8|, \verb|LC_NUMERIC=C|, \verb|LC_TIME=en_US.UTF-8|, \verb|LC_COLLATE=en_US.UTF-8|, \verb|LC_MONETARY=en_US.UTF-8|, \verb|LC_MESSAGES=en_US.UTF-8|, \verb|LC_PAPER=en_US.UTF-8|, \verb|LC_NAME=C|, \verb|LC_ADDRESS=C|, \verb|LC_TELEPHONE=C|, \verb|LC_MEASUREMENT=en_US.UTF-8|, \verb|LC_IDENTIFICATION=C|
  \item Time zone: \verb|Europe/Brussels|
  \item TZcode source: \verb|system (glibc)|
  \item Running under: \verb|Manjaro Linux|
  \item Matrix products: default
  \item BLAS:   \verb|/usr/lib/libblas.so.3.12.0|
  \item LAPACK: \verb|/usr/lib/liblapack.so.3.12.0|
  \item Base packages: base, datasets, graphics, grDevices, methods,
    stats, stats4, utils
  \item Other packages: Biobase~2.62.0, BiocGenerics~0.48.1,
    dplyr~1.1.4, GenomeInfoDb~1.38.1, GenomicRanges~1.54.1,
    ggplot2~3.4.4, IRanges~2.36.0, kableExtra~1.3.4, limma~3.58.1,
    MatrixGenerics~1.14.0, matrixStats~1.2.0,
    MultiAssayExperiment~1.28.0, patchwork~1.1.3, QFeatures~1.12.0,
    rmarkdown~2.25, S4Vectors~0.40.2, scater~1.30.1, scp~1.12.0,
    scuttle~1.12.0, SingleCellExperiment~1.24.0,
    SummarizedExperiment~1.32.0
  \item Loaded via a namespace (and not attached): abind~1.4-5,
    AnnotationFilter~1.26.0, beachmat~2.18.0, beeswarm~0.4.0,
    BiocBaseUtils~1.4.0, BiocNeighbors~1.20.0, BiocParallel~1.36.0,
    BiocSingular~1.18.0, bitops~1.0-7, bookdown~0.37, cli~3.6.2,
    clue~0.3-65, cluster~2.1.4, codetools~0.2-19, colorspace~2.1-0,
    compiler~4.3.1, cowplot~1.1.1, crayon~1.5.2, DelayedArray~0.28.0,
    DelayedMatrixStats~1.24.0, digest~0.6.33, evaluate~0.23,
    fansi~1.0.6, farver~2.1.1, fastmap~1.1.1, generics~0.1.3,
    GenomeInfoDbData~1.2.11, ggbeeswarm~0.7.2, ggrepel~0.9.4,
    glue~1.6.2, grid~4.3.1, gridExtra~2.3, gtable~0.3.4, highr~0.10,
    htmltools~0.5.7, httr~1.4.7, igraph~1.6.0, irlba~2.3.5.1,
    knitr~1.45, labeling~0.4.3, lattice~0.22-5, lazyeval~0.2.2,
    lifecycle~1.0.4, magrittr~2.0.3, MASS~7.3-60, Matrix~1.6-4,
    MsCoreUtils~1.15.1, munsell~0.5.0, parallel~4.3.1, pillar~1.9.0,
    pkgconfig~2.0.3, ProtGenerics~1.34.0, R6~2.5.1, Rcpp~1.0.11,
    RCurl~1.98-1.13, rlang~1.1.2, rstudioapi~0.15.0, rsvd~1.0.5,
    rvest~1.0.3, S4Arrays~1.2.0, ScaledMatrix~1.10.0, scales~1.3.0,
    SparseArray~1.2.2, sparseMatrixStats~1.14.0, statmod~1.5.0,
    stringi~1.8.3, stringr~1.5.1, svglite~2.1.3, systemfonts~1.0.5,
    tibble~3.2.1, tidyselect~1.2.0, tools~4.3.1, utf8~1.2.4,
    vctrs~0.6.5, vipor~0.4.5, viridis~0.6.4, viridisLite~0.4.2,
    webshot~0.5.5, withr~2.5.2, xfun~0.41, xml2~1.3.6, XVector~0.42.0,
    yaml~2.3.8, zlibbioc~1.48.0
\end{itemize}

\textbf{Acknowledgements}: This work was funded by a research fellowship of
the Fonds National de la Recherche Scientifique (FNRS) to CV. SG is a
FRIA grantee of the FNRS. CK is supported by the Walloon Region SPW
through the Win2Wal grant 2010126 (ChipOmics). The TimsTOF SCP was
funded thanks to the Foundation against Cancer (agreement 2022-040A).

\hypertarget{bibliography}{%
\section*{Bibliography}\label{bibliography}}
\addcontentsline{toc}{section}{Bibliography}

\hypertarget{refs}{}
\begin{CSLReferences}{0}{0}
\leavevmode\vadjust pre{\hypertarget{ref-leduc_exploring_2022}{}}%
\CSLLeftMargin{(1) }%
\CSLRightInline{Leduc, A.; Huffman, R. G.; Cantlon, J.; Khan, S.; Slavov, N. Exploring Functional Protein Covariation Across Single Cells Using {nPOP}. \emph{Genome Biology} \textbf{2022}, \emph{23} (1), 261. \url{https://doi.org/10.1186/s13059-022-02817-5}.}

\leavevmode\vadjust pre{\hypertarget{ref-derks_increasing_2023}{}}%
\CSLLeftMargin{(2) }%
\CSLRightInline{Derks, J.; Leduc, A.; Wallmann, G.; Huffman, R. G.; Willetts, M.; Khan, S.; Specht, H.; Ralser, M.; Demichev, V.; Slavov, N. Increasing the Throughput of Sensitive Proteomics by {plexDIA}. \emph{Nature Biotechnology} \textbf{2023}, \emph{41} (1), 50--59. \url{https://doi.org/10.1038/s41587-022-01389-w}.}

\leavevmode\vadjust pre{\hypertarget{ref-matzinger_robust_2023}{}}%
\CSLLeftMargin{(3) }%
\CSLRightInline{Matzinger, M.; Müller, E.; Dürnberger, G.; Pichler, P.; Mechtler, K. Robust and {Easy}-to-{Use} {One}-{Pot} {Workflow} for {Label}-{Free} {Single}-{Cell} {Proteomics}. \emph{Analytical Chemistry} \textbf{2023}, \emph{95} (9), 4435--4445. \url{https://doi.org/10.1021/acs.analchem.2c05022}.}

\leavevmode\vadjust pre{\hypertarget{ref-slavov_learning_2022}{}}%
\CSLLeftMargin{(4) }%
\CSLRightInline{Slavov, N. Learning from Natural Variation Across the Proteomes of Single Cells. \emph{PLOS Biology} \textbf{2022}, \emph{20} (1), e3001512. \url{https://doi.org/10.1371/journal.pbio.3001512}.}

\leavevmode\vadjust pre{\hypertarget{ref-vanderaa_current_2023}{}}%
\CSLLeftMargin{(5) }%
\CSLRightInline{Vanderaa, C.; Gatto, L. The {Current} {State} of {Single}-{Cell} {Proteomics} {Data} {Analysis}. \emph{Current Protocols} \textbf{2023}, \emph{3} (1), e658. \url{https://doi.org/10.1002/cpz1.658}.}

\leavevmode\vadjust pre{\hypertarget{ref-vanderaa_replication_2021}{}}%
\CSLLeftMargin{(6) }%
\CSLRightInline{Vanderaa, C.; Gatto, L. Replication of Single-Cell Proteomics Data Reveals Important Computational Challenges. \emph{Expert Review of Proteomics} \textbf{2021}, \emph{18} (10), 835--843. \url{https://doi.org/10.1080/14789450.2021.1988571}.}

\leavevmode\vadjust pre{\hypertarget{ref-huber_orchestrating_2015}{}}%
\CSLLeftMargin{(7) }%
\CSLRightInline{Huber, W.; Carey, V. J.; Gentleman, R.; Anders, S.; Carlson, M.; Carvalho, B. S.; Bravo, H. C.; Davis, S.; Gatto, L.; Girke, T.; Gottardo, R.; Hahne, F.; Hansen, K. D.; Irizarry, R. A.; Lawrence, M.; Love, M. I.; MacDonald, J.; Obenchain, V.; Oleś, A. K.; Pagès, H.; Reyes, A.; Shannon, P.; Smyth, G. K.; Tenenbaum, D.; Waldron, L.; Morgan, M. Orchestrating High-Throughput Genomic Analysis with {Bioconductor}. \emph{Nature Methods} \textbf{2015}, \emph{12} (2), 115--121. \url{https://doi.org/10.1038/nmeth.3252}.}

\leavevmode\vadjust pre{\hypertarget{ref-R-SingleCellExperiment}{}}%
\CSLLeftMargin{(8) }%
\CSLRightInline{Lun, A.; Risso, D. \emph{SingleCellExperiment: S4 Classes for Single Cell Data}; 2023. \url{https://doi.org/10.18129/B9.bioc.SingleCellExperiment}.}

\leavevmode\vadjust pre{\hypertarget{ref-SingleCellExperiment2020}{}}%
\CSLLeftMargin{(9) }%
\CSLRightInline{Amezquita, R.; Lun, A.; Becht, E.; Carey, V.; Carpp, L.; Geistlinger, L.; Marini, F.; Rue-Albrecht, K.; Risso, D.; Soneson, C.; Waldron, L.; Pages, H.; Smith, M.; Huber, W.; Morgan, M.; Gottardo, R.; Hicks, S. \href{https://www.nature.com/articles/s41592-019-0654-x}{Orchestrating Single-Cell Analysis with Bioconductor}. \emph{Nature Methods} \textbf{2020}, \emph{17}, 137--145.}

\leavevmode\vadjust pre{\hypertarget{ref-tian_benchmarking_2019}{}}%
\CSLLeftMargin{(10) }%
\CSLRightInline{Tian, L.; Dong, X.; Freytag, S.; Lê Cao, K.-A.; Su, S.; JalalAbadi, A.; Amann-Zalcenstein, D.; Weber, T. S.; Seidi, A.; Jabbari, J. S.; Naik, S. H.; Ritchie, M. E. Benchmarking Single Cell {RNA}-Sequencing Analysis Pipelines Using Mixture Control Experiments. \emph{Nature Methods} \textbf{2019}, \emph{16} (6), 479--487. \url{https://doi.org/10.1038/s41592-019-0425-8}.}

\leavevmode\vadjust pre{\hypertarget{ref-Mereu2020-vk}{}}%
\CSLLeftMargin{(11) }%
\CSLRightInline{Mereu, E.; Lafzi, A.; Moutinho, C.; Ziegenhain, C.; McCarthy, D. J.; Álvarez-Varela, A.; Batlle, E.; Sagar; Grün, D.; Lau, J. K.; Boutet, S. C.; Sanada, C.; Ooi, A.; Jones, R. C.; Kaihara, K.; Brampton, C.; Talaga, Y.; Sasagawa, Y.; Tanaka, K.; Hayashi, T.; Braeuning, C.; Fischer, C.; Sauer, S.; Trefzer, T.; Conrad, C.; Adiconis, X.; Nguyen, L. T.; Regev, A.; Levin, J. Z.; Parekh, S.; Janjic, A.; Wange, L. E.; Bagnoli, J. W.; Enard, W.; Gut, M.; Sandberg, R.; Nikaido, I.; Gut, I.; Stegle, O.; Heyn, H. Benchmarking Single-Cell {RNA-sequencing} Protocols for Cell Atlas Projects. \emph{Nat. Biotechnol.} \textbf{2020}, \emph{38} (6), 747--755.}

\leavevmode\vadjust pre{\hypertarget{ref-Hadley_dplyr_2023}{}}%
\CSLLeftMargin{(12) }%
\CSLRightInline{Wickham, H.; François, R.; Henry, L.; Müller, K.; Vaughan, D. \emph{\href{https://CRAN.R-project.org/package=dplyr}{Dplyr: A Grammar of Data Manipulation}}; 2023.}

\leavevmode\vadjust pre{\hypertarget{ref-Hadley_gpglot2_2016}{}}%
\CSLLeftMargin{(13) }%
\CSLRightInline{Wickham, H. \emph{\href{https://ggplot2.tidyverse.org}{Ggplot2: Elegant Graphics for Data Analysis}}; Springer-Verlag New York, 2016.}

\leavevmode\vadjust pre{\hypertarget{ref-Ritchie_limma_2015}{}}%
\CSLLeftMargin{(14) }%
\CSLRightInline{Ritchie, M. E.; Phipson, B.; Wu, D.; Hu, Y.; Law, C. W.; Shi, W.; Smyth, G. K. {limma} Powers Differential Expression Analyses for {RNA}-Sequencing and Microarray Studies. \emph{Nucleic Acids Research} \textbf{2015}, \emph{43} (7), e47. \url{https://doi.org/10.1093/nar/gkv007}.}

\leavevmode\vadjust pre{\hypertarget{ref-McCarthy_scater_2017}{}}%
\CSLLeftMargin{(15) }%
\CSLRightInline{McCarthy, D. J.; Campbell, K. R.; Lun, A. T. L.; Willis, Q. F. Scater: Pre-Processing, Quality Control, Normalisation and Visualisation of Single-Cell {R}{N}{A}-Seq Data in {R}. \emph{Bioinformatics} \textbf{2017}, \emph{33}, 1179--1186. \url{https://doi.org/10.1093/bioinformatics/btw777}.}

\leavevmode\vadjust pre{\hypertarget{ref-specht_single-cell_2021}{}}%
\CSLLeftMargin{(16) }%
\CSLRightInline{Specht, H.; Emmott, E.; Petelski, A. A.; Huffman, R. G.; Perlman, D. H.; Serra, M.; Kharchenko, P.; Koller, A.; Slavov, N. Single-Cell Proteomic and Transcriptomic Analysis of Macrophage Heterogeneity Using {SCoPE2}. \emph{Genome Biology} \textbf{2021}, \emph{22} (1), 50. \url{https://doi.org/10.1186/s13059-021-02267-5}.}

\leavevmode\vadjust pre{\hypertarget{ref-adusumilli_data_2017}{}}%
\CSLLeftMargin{(17) }%
\CSLRightInline{Adusumilli, R.; Mallick, P. Data {Conversion} with {ProteoWizard} {msConvert}. In \emph{Proteomics: {Methods} and {Protocols}}; Comai, L., Katz, J. E., Mallick, P., Eds.; Methods in {Molecular} {Biology}; Springer: New York, NY, 2017; pp 339--368. \url{https://doi.org/10.1007/978-1-4939-6747-6_23}.}

\leavevmode\vadjust pre{\hypertarget{ref-lazear_sage_2023}{}}%
\CSLLeftMargin{(18) }%
\CSLRightInline{Lazear, M. R. Sage: {An} {Open}-{Source} {Tool} for {Fast} {Proteomics} {Searching} and {Quantification} at {Scale}. \emph{Journal of Proteome Research} \textbf{2023}. \url{https://doi.org/10.1021/acs.jproteome.3c00486}.}

\leavevmode\vadjust pre{\hypertarget{ref-zenododata}{}}%
\CSLLeftMargin{(19) }%
\CSLRightInline{Grégoire, S.; Vanderaa, C.; Pyr dit Ruys, S.; Kune, C.; Mazzucchelli, G.; Vertommen, D.; Gatto, L. Data Accompanying "Standardised Workflow for Mass Spectrometry-Based Single-Cell Proteomics Data Analysis Using the Scp Package". \emph{Zenodo} \textbf{2023}. \url{https://doi.org/10.5281/zenodo.8417228}.}

\leavevmode\vadjust pre{\hypertarget{ref-Vizcaino2014}{}}%
\CSLLeftMargin{(20) }%
\CSLRightInline{Vizcaíno, J. A.; Deutsch, E. W.; Wang, R.; Csordas, A.; Reisinger, F.; Ríos, D.; Dianes, J. A.; Sun, Z.; Farrah, T.; Bandeira, N.; Binz, P.-A.; Xenarios, I.; Eisenacher, M.; Mayer, G.; Gatto, L.; Campos, A.; Chalkley, R. J.; Kraus, H.-J.; Albar, J. P.; Martinez-Bartolomé, S.; Apweiler, R.; Omenn, G. S.; Martens, L.; Jones, A. R.; Hermjakob, H. {ProteomeXchange} Provides Globally Coordinated Proteomics Data Submission and Dissemination. \emph{Nat. Biotechnol.} \textbf{2014}, \emph{32} (3), 223--226.}

\leavevmode\vadjust pre{\hypertarget{ref-cox_maxquant_2008}{}}%
\CSLLeftMargin{(21) }%
\CSLRightInline{Cox, J.; Mann, M. {MaxQuant} Enables High Peptide Identification Rates, Individualized p.p.b.-Range Mass Accuracies and Proteome-Wide Protein Quantification. \emph{Nature Biotechnology} \textbf{2008}, \emph{26} (12), 1367--1372. \url{https://doi.org/10.1038/nbt.1511}.}

\leavevmode\vadjust pre{\hypertarget{ref-kong_msfragger_2017}{}}%
\CSLLeftMargin{(22) }%
\CSLRightInline{Kong, A. T.; Leprevost, F. V.; Avtonomov, D. M.; Mellacheruvu, D.; Nesvizhskii, A. I. {MSFragger}: Ultrafast and Comprehensive Peptide Identification in Mass Spectrometry--Based Proteomics. \emph{Nature Methods} \textbf{2017}, \emph{14} (5), 513--520. \url{https://doi.org/10.1038/nmeth.4256}.}

\leavevmode\vadjust pre{\hypertarget{ref-gatto_initial_2023}{}}%
\CSLLeftMargin{(23) }%
\CSLRightInline{Gatto, L.; Aebersold, R.; Cox, J.; Demichev, V.; Derks, J.; Emmott, E.; Franks, A. M.; Ivanov, A. R.; Kelly, R. T.; Khoury, L.; Leduc, A.; MacCoss, M. J.; Nemes, P.; Perlman, D. H.; Petelski, A. A.; Rose, C. M.; Schoof, E. M.; Van Eyk, J.; Vanderaa, C.; Yates, J. R.; Slavov, N. Initial Recommendations for Performing, Benchmarking and Reporting Single-Cell Proteomics Experiments. \emph{Nature Methods} \textbf{2023}, \emph{20} (3), 375--386. \url{https://doi.org/10.1038/s41592-023-01785-3}.}

\leavevmode\vadjust pre{\hypertarget{ref-Kong2022-wp}{}}%
\CSLLeftMargin{(24) }%
\CSLRightInline{Kong, W.; Hui, H. W. H.; Peng, H.; Goh, W. W. B. Dealing with Missing Values in Proteomics Data. \emph{Proteomics} \textbf{2022}, \emph{22} (23-24), e2200092.}

\leavevmode\vadjust pre{\hypertarget{ref-cuklina_diagnostics_2021}{}}%
\CSLLeftMargin{(25) }%
\CSLRightInline{Čuklina, J.; Lee, C. H.; Williams, E. G.; Sajic, T.; Collins, B. C.; Rodríguez Martínez, M.; Sharma, V. S.; Wendt, F.; Goetze, S.; Keele, G. R.; Wollscheid, B.; Aebersold, R.; Pedrioli, P. G. A. Diagnostics and Correction of Batch Effects in Large‐scale Proteomic Studies: A Tutorial. \emph{Molecular Systems Biology} \textbf{2021}, \emph{17} (8), e10240. \url{https://doi.org/10.15252/msb.202110240}.}

\leavevmode\vadjust pre{\hypertarget{ref-obrien_effects_2018}{}}%
\CSLLeftMargin{(26) }%
\CSLRightInline{O'Brien, J. J.; Gunawardena, H. P.; Paulo, J. A.; Chen, X.; Ibrahim, J. G.; Gygi, S. P.; Qaqish, B. F. The Effects of Nonignorable Missing Data on Label-Free Mass Spectrometry Proteomics Experiments. \emph{The annals of applied statistics} \textbf{2018}, \emph{12} (4), 2075--2095. \url{https://doi.org/10.1214/18-AOAS1144}.}

\leavevmode\vadjust pre{\hypertarget{ref-goeminne_msqrob_2020}{}}%
\CSLLeftMargin{(27) }%
\CSLRightInline{Goeminne, L. J. E.; Sticker, A.; Martens, L.; Gevaert, K.; Clement, L. {MSqRob} {Takes} the {Missing} {Hurdle}: {Uniting} {Intensity}- and {Count}-{Based} {Proteomics}. \emph{Analytical Chemistry} \textbf{2020}, \emph{92} (9), 6278--6287. \url{https://doi.org/10.1021/acs.analchem.9b04375}.}

\leavevmode\vadjust pre{\hypertarget{ref-vanderaa_revisiting_2023}{}}%
\CSLLeftMargin{(28) }%
\CSLRightInline{Vanderaa, C.; Gatto, L. Revisiting the {Thorny} {Issue} of {Missing} {Values} in {Single}-{Cell} {Proteomics}. \emph{Journal of Proteome Research} \textbf{2023}, \emph{22} (9), 2775--2784. \url{https://doi.org/10.1021/acs.jproteome.3c00227}.}

\leavevmode\vadjust pre{\hypertarget{ref-schoof_quantitative_2021}{}}%
\CSLLeftMargin{(29) }%
\CSLRightInline{Schoof, E. M.; Furtwängler, B.; Üresin, N.; Rapin, N.; Savickas, S.; Gentil, C.; Lechman, E.; Keller, U. auf dem; Dick, J. E.; Porse, B. T. Quantitative Single-Cell Proteomics as a Tool to Characterize Cellular Hierarchies. \emph{Nature Communications} \textbf{2021}, \emph{12}, 3341. \url{https://doi.org/10.1038/s41467-021-23667-y}.}

\leavevmode\vadjust pre{\hypertarget{ref-petrosius_exploration_2023}{}}%
\CSLLeftMargin{(30) }%
\CSLRightInline{Petrosius, V.; Aragon-Fernandez, P.; Üresin, N.; Kovacs, G.; Phlairaharn, T.; Furtwängler, B.; Op De Beeck, J.; Skovbakke, S. L.; Goletz, S.; Thomsen, S. F.; Keller, U. auf dem; Natarajan, K. N.; Porse, B. T.; Schoof, E. M. Exploration of Cell State Heterogeneity Using Single-Cell Proteomics Through Sensitivity-Tailored Data-Independent Acquisition. \emph{Nature Communications} \textbf{2023}, \emph{14}, 5910. \url{https://doi.org/10.1038/s41467-023-41602-1}.}

\leavevmode\vadjust pre{\hypertarget{ref-Lun_scran_2016}{}}%
\CSLLeftMargin{(31) }%
\CSLRightInline{Lun, A. T. L.; McCarthy, D. J.; Marioni, J. C. A Step-by-Step Workflow for Low-Level Analysis of Single-Cell RNA-Seq Data with Bioconductor. \emph{F1000Res.} \textbf{2016}, \emph{5}, 2122. \url{https://doi.org/10.12688/f1000research.9501.2}.}

\leavevmode\vadjust pre{\hypertarget{ref-liang_fully_2021}{}}%
\CSLLeftMargin{(32) }%
\CSLRightInline{Liang, Y.; Acor, H.; McCown, M. A.; Nwosu, A. J.; Boekweg, H.; Axtell, N. B.; Truong, T.; Cong, Y.; Payne, S. H.; Kelly, R. T. Fully {Automated} {Sample} {Processing} and {Analysis} {Workflow} for {Low}-{Input} {Proteome} {Profiling}. \emph{Analytical Chemistry} \textbf{2021}, \emph{93} (3), 1658--1666. \url{https://doi.org/10.1021/acs.analchem.0c04240}.}

\leavevmode\vadjust pre{\hypertarget{ref-brunner_ultrahigh_2022}{}}%
\CSLLeftMargin{(33) }%
\CSLRightInline{Brunner, A.; Thielert, M.; Vasilopoulou, C.; Ammar, C.; Coscia, F.; Mund, A.; Hoerning, O. B.; Bache, N.; Apalategui, A.; Lubeck, M.; Richter, S.; Fischer, D. S.; Raether, O.; Park, M. A.; Meier, F.; Theis, F. J.; Mann, M. Ultra‐high Sensitivity Mass Spectrometry Quantifies Single‐cell Proteome Changes Upon Perturbation. \emph{Molecular Systems Biology} \textbf{2022}, \emph{18} (3), e10798. \url{https://doi.org/10.15252/msb.202110798}.}

\leavevmode\vadjust pre{\hypertarget{ref-Wu_clusterProfiler_2021}{}}%
\CSLLeftMargin{(34) }%
\CSLRightInline{Wu, T.; Hu, E.; Xu, S.; Chen, M.; Guo, P.; Dai, Z.; Feng, T.; Zhou, L.; Tang, W.; Zhan, L.; Fu, xiaochong; Liu, S.; Bo, X.; Yu, G. clusterProfiler 4.0: A Universal Enrichment Tool for Interpreting Omics Data. \emph{The Innovation} \textbf{2021}, \emph{2} (3), 100141. \url{https://doi.org/10.1016/j.xinn.2021.100141}.}

\leavevmode\vadjust pre{\hypertarget{ref-Yu_clusterProfiler_2012}{}}%
\CSLLeftMargin{(35) }%
\CSLRightInline{Yu, G.; Wang, L.-G.; Han, Y.; He, Q.-Y. clusterProfiler: An r Package for Comparing Biological Themes Among Gene Clusters. \emph{OMICS: A Journal of Integrative Biology} \textbf{2012}, \emph{16} (5), 284--287. \url{https://doi.org/10.1089/omi.2011.0118}.}

\leavevmode\vadjust pre{\hypertarget{ref-Angerer_destiny_2015}{}}%
\CSLLeftMargin{(36) }%
\CSLRightInline{Angerer, P.; Haghverdi, L.; Büttner, M.; Theis, F.; Marr, C.; Büttner, F. Destiny: Diffusion Maps for Large-Scale Single-Cell Data in r. \emph{Bioinformatics} \textbf{2015}, \emph{32} (8), 1243. \url{https://doi.org/10.1093/bioinformatics/btv715}.}

\leavevmode\vadjust pre{\hypertarget{ref-zhu_single-cell_2019}{}}%
\CSLLeftMargin{(37) }%
\CSLRightInline{Zhu, Y.; Scheibinger, M.; Ellwanger, D. C.; Krey, J. F.; Choi, D.; Kelly, R. T.; Heller, S.; Barr-Gillespie, P. G. Single-Cell Proteomics Reveals Changes in Expression During Hair-Cell Development. \emph{eLife} \textbf{2019}, \emph{8}, e50777. \url{https://doi.org/10.7554/eLife.50777}.}

\leavevmode\vadjust pre{\hypertarget{ref-Ellwanger_CellTrails_2018}{}}%
\CSLLeftMargin{(38) }%
\CSLRightInline{Ellwanger, D. C.; Scheibinger, M.; Dumont, R. A.; Barr-Gillespie, P. G.; Heller, S. Transcriptional Dynamics of Hair-Bundle Morphogenesis Revealed with CellTrails. \emph{Cell Reports} \textbf{2018}, \emph{23(10)}, 2901--2914. \url{https://doi.org/10.1016/j.celrep.2018.05.002}.}

\leavevmode\vadjust pre{\hypertarget{ref-sticker_robust_2020}{}}%
\CSLLeftMargin{(39) }%
\CSLRightInline{Sticker, A.; Goeminne, L.; Martens, L.; Clement, L. Robust {Summarization} and {Inference} in {Proteome}-Wide {Label}-Free {Quantification}. \emph{Molecular \& Cellular Proteomics} \textbf{2020}, \emph{19} (7), 1209--1219. \url{https://doi.org/10.1074/mcp.RA119.001624}.}

\leavevmode\vadjust pre{\hypertarget{ref-tu_systematic_2014}{}}%
\CSLLeftMargin{(40) }%
\CSLRightInline{Tu, C.; Li, J.; Sheng, Q.; Zhang, M.; Qu, J. Systematic {Assessment} of {Survey} {Scan} and {MS2}-{Based} {Abundance} {Strategies} for {Label}-{Free} {Quantitative} {Proteomics} {Using} {High}-{Resolution} {MS} {Data}. \emph{Journal of Proteome Research} \textbf{2014}, \emph{13} (4), 2069--2079. \url{https://doi.org/10.1021/pr401206m}.}

\leavevmode\vadjust pre{\hypertarget{ref-lazar_accounting_2016}{}}%
\CSLLeftMargin{(41) }%
\CSLRightInline{Lazar, C.; Gatto, L.; Ferro, M.; Bruley, C.; Burger, T. Accounting for the {Multiple} {Natures} of {Missing} {Values} in {Label}-{Free} {Quantitative} {Proteomics} {Data} {Sets} to {Compare} {Imputation} {Strategies}. \emph{Journal of Proteome Research} \textbf{2016}, \emph{15} (4), 1116--1125. \url{https://doi.org/10.1021/acs.jproteome.5b00981}.}

\leavevmode\vadjust pre{\hypertarget{ref-rainer_modular_2022}{}}%
\CSLLeftMargin{(42) }%
\CSLRightInline{Rainer, J.; Vicini, A.; Salzer, L.; Stanstrup, J.; Badia, J. M.; Neumann, S.; Stravs, M. A.; Verri Hernandes, V.; Gatto, L.; Gibb, S.; Witting, M. A Modular and Expandable Ecosystem for Metabolomics Data Annotation in r. \emph{Metabolites} \textbf{2022}, \emph{12}, 173. \url{https://doi.org/10.3390/metabo12020173}.}

\end{CSLReferences}

\end{document}